\documentclass[apj,twocolumn,twocolappendix,numberedappendix]{openjournal}
\usepackage[dvipsnames]{xcolor}
\usepackage{newtxtext,newtxmath,enumitem,multirow}
\usepackage[utf8]{inputenc}
\usepackage[T1]{fontenc}
\usepackage[breaklinks,colorlinks,allcolors=blue,citecolor=blue,urlcolor=blue]{hyperref}
\usepackage{orcidlink}
\usepackage{graphicx}	% Including figure files
\usepackage{amsmath}	% Advanced maths commands
\usepackage{amssymb}	% Extra maths symbols
\usepackage{pifont}
\usepackage{mathrsfs}
\usepackage{blindtext}
\usepackage[dvipsnames]{xcolor}

\pdfoutput=1

\newcommand{\LCDM}{$\Lambda$CDM\ }
\newcommand{\lya}{Ly$\alpha$\ }

\newcommand{\kpc}{\: \mathrm{kpc}}

\newcommand{\ckpc}{\: \mathrm{ckpc}}
\newcommand{\cMpc}{\: \mathrm{cMpc}}
\newcommand{\hMpc}{\: h^{-1} \, \mathrm{cMpc}}
\newcommand{\Msun}{\: \mathrm{M}_{\odot}}
\newcommand{\K}{\: \mathrm{K}}
\newcommand{\simba}{\textsc{Simba}}
\newcommand{\mufasa}{\textsc{Mufasa}}

\newcommand{\cmark}{\ding{51}}%

\shorttitle{A universal fitting formula for gas density profiles}
\shortauthors{D. Sorini et al.}

\begin{document}
\title[The impact of feedback on the evolution of gas density profiles from galaxies to clusters: a universal fitting formula from the Simba suite of simulations]{The impact of feedback on the evolution of gas density profiles from galaxies to clusters: a universal fitting formula from the Simba suite of simulations}
\author{Daniele Sorini\,\orcidlink{0000-0001-5252-9561}$^{1,\star}$}
\author{Sownak Bose\,\orcidlink{0000-0002-0974-5266}$^{1}$}
\author{Romeel Dav\'e\,\orcidlink{0000-0003-2842-9434}$^{2,3}$}
\author{Daniel Angl\'es-Alc\'azar\,\orcidlink{0000-0001-5769-4945}$^{4}$}
% List of institutions
\affiliation{$^{1}$Institute for Computational Cosmology, Department of Physics, Durham University, South Road, Durham, DH1 3LE, United Kingdom}
\affiliation{$^{2}$Institute for Astronomy, University of Edinburgh, Royal Observatory, Blackford Hill, Edinburgh, EH9 3HJ, United Kingdom}
\affiliation{$^{3}$University of the Western Cape, Bellville, Cape Town 7535, South Africa}
\affiliation{$^{4}$Department of Physics, University of Connecticut, 196 Auditorium Road, U-3046, Storrs, CT 06269-3046, USA}
\thanks{$^\star$\href{mailto:daniele.sorini@durham.ac.uk}{daniele.sorini@durham.ac.uk}}

\begin{abstract}
The radial distribution of gas within galactic haloes is connected to the star formation rate and the nature of baryon-driven feedback processes. Using six variants of the hydrodynamic simulation \simba, we study the impact of different stellar/AGN feedback prescriptions on the gas density profiles of haloes in the mass range $10^{11} \, \mathrm{M}_{\odot} < M_{\rm 200c} < 10^{14} \Msun$ and redshift interval $0<z<4$. We find that the radial profiles are well represented by a power law and that, for a fixed total halo mass, the slope and amplitude of such power law are generally weakly dependent on redshift. Once AGN-driven jets are activated in the simulation, the gas density profile of haloes with $M_{\rm 200c} \gtrsim 10^{13} \, \rm M_{\odot}$ declines more gently with radial distance. We argue that this distinctive feature could be exploited with current observations to discriminate amongst the predictions of the different feedback models. We introduce a universal fitting formula for the slope and amplitude of the gas density profile as a function of mass and redshift. The best-fit functions are suitable for all feedback variants considered, and their predictions are in excellent agreement with the numerical results. We provide the values of all fit parameters, making our fitting formula a versatile tool to mimic the effect of \simba\ feedback models onto N-body simulations and semi-analytical models of galaxy formation. Our results can also aid observational estimates of the gas mass within haloes that assume a specific slope for the underlying gas density profile.
\end{abstract}


\maketitle
%%%%%%%%%%%%%%%%%%%%%%%%%%%%%%%%%%%%%%%%%%%%%%%%%%

%%%%%%%%%%%%%%%%% BODY OF PAPER %%%%%%%%%%%%%%%%%%

%------------------------------------------------------
\section{Introduction}
\label{sec:intro}
%------------------------------------------------------

%\blindtext

Understanding the distribution of gas within galactic halos is pivotal to unravelling the complexities of galaxy formation and evolution. The efficiency of star formation within galaxies depends on the amount of gas and its thermal state. These are directly affected by complex astrophysical feedback processes such as supernovae explosions, stellar winds or outflows driven by active galactic nuclei (AGN), which are not yet fully understood but are thought to play a critical role in shaping the properties of the galaxy population through cosmic time \citep[see e.g.][for recent reviews]{feedback_review, Crain_2023}. Thus, observing and modelling the radial distribution of different gaseous phases within galaxies and their surrounding circumgalactic medium (CGM) has been the focus of a large body of research.

Several observational campaigns exploited the \lya absorption line from background quasars to probe the CGM of foreground galaxies \citep{Steidel_2010, Rudie_2012_CGM, Rudie_2013, Turner_2014}, damped \lya systems \citep{Rubin_2015} and quasars \citep[e.g][]{QPQ2, Prochaska_2013} at redshift $2\lesssim z \lesssim 3$. At lower redshift, metal absorbers have been utilised to study the spatial extent of the cool CGM of quasars \citep{Farina_2011, Farina_2013, Farina_2014, Johnson_2015, Johnson_2015_environment, Johnson_2016}. These studies are complemented by observations of \lya and metal emission lines around $z\gtrsim 2$ quasars \cite[e.g.][]{Arrigoni-Battaia_2015}. Integral-field spectroscopic instruments such as MUSE \citep{MUSE} have enabled further studies of the distribution of cool gas around $2 \lesssim z \lesssim 3$ quasars \citep[e.g.][]{Arrigoni-Battaia_2019, Arrigoni-Battaia_2023} and even higher-redshift ($z<4.5$) galaxies \citep[e.g.][]{Galbiati_2023, Galbiati_2024}. Recently, the radial surface brightness profiles of hot gas around low-redshift galaxies and clusters have been mapped \citep{Lyskova_2023, Zhang_2024} thanks to the eROSITA X-ray observatory \citep{eROSITA}. Observations of the thermal and kinetic Sunyaev-Zeldovich effects \citep{SZ_1970, SZ_1972} have enabled constraining thermodynamical properties of the CGM around galaxies, groups, and of the intracluster medium \citep{Amodeo_2021}. Thus, the combination of a large amount of data from different wavelength bands enables us to trace the distribution of different gaseous phases around a wide mass range of objects, over a considerable redshift interval. This wealth of data is invaluable for constraining models for the build up of galaxies in a cosmological context, which often rely on specific assumptions on the gas density profiles within haloes.

Widely used semi-analytical models of galaxy formation adopt a spherically symmetric density profile declining as the inverse square of the distance from the halo centre \citep{Cole_1994, Kauffmann_1999, HS03, Lacey_2016, Pandya_2020}. While this assumption was later validated by cosmological simulations, at least in the outskirts of haloes \citep{Stinson_2015}, other choices appeared in the literature as well. For instance, some fully analytical frameworks for cosmic star formation history or models of the CGM structure prefer a generic power law \citep{HS03, SP21, Pandya_2023}. Other authors \citep{Matthews_2017, Prochaska_2019_FRB}  proposed a modification of the so-called `NFW profile', which is well known to accurately describe dark matter density profiles in N-body simulations \citep{NFW}. Another recurrent functional shape for the gas density profile is the `beta model', which is empirically motivated by both observations \citep[e.g.][]{Zhang_2024} and simulations \citep{Suginohara_1998}. Different analytical solutions for the gas density profiles have been derived from first principles \citep{Capelo_2010, Stern_2016, Stern_2019}, providing further insight into the connection between astrophysical processes and the radial distribution of gas within haloes.

Full cosmological hydrodynamic simulations can provide a more accurate description of the gas distribution within and around galaxies, without invoking simplifying assumptions such as spherical symmetry. However, depending on the exact sub-grid implementation of stellar and AGN feedback processes, the predictions may vary considerably from code to code. For example, \cite{Pallottini_2014} showed that simulations based on the \textsc{RAMSES} adaptive mesh-refinement code \citep{Ramses} produce neutral hydrogen density profiles that are in agreement with \lya absorption measurements in the CGM of $z\sim 2$ galaxies \citep{Steidel_2010}. The Nyx \citep{Almgren_2013, Lukic_2015} and Illustris \citep{Illustris_V2014} simulation provide a good match to similar observations \citep{Turner_2014}, except in the region within $10 \kpc$ from galaxies, despite exhibiting rather different gas density and temperature profiles \citep{Sorini_2018}. On the other hand, the Illustris simulation predicts a hot gas mass distribution that is broadly consistent with observations of the X-ray coronal emission from low-mass spirals \citep{Bogdan_2015}. 

More recently, the CAMELS suite of cosmological simulations \citep{CAMELS} has been used in conjunction with thermal SZ data from the Atacama Cosmology Telescope (ACT; \citealt{Madhavacheril_2020}) and lensing measurements from the Dark Energy Survey (DES; \citealt{Amon_2022, Secco_2022}) to constrain the effect of feedback prescriptions on the matter power spectrum \citep{Moser_2022, Pandey_2023}. Other works with the CAMELS suite explicitly highlighted the dependence on the parameters of different feedback models of the the gas distribution across various scales \cite{Gebhardt_2024}, the halo baryon mass \cite{Delgado_2023} fraction and the gas power spectrum \cite{Ni_2023}. 

Further insights were offered by zoom-in numerical campaigns. The FIRE-2  simulations \citep{FIRE-2} showed that the inclusion of cosmic rays may significantly affect the gas density profiles of galaxies more massive than the Milky Way, without appreciably altering the temperature. Different implementations of the code underlying the \textsc{The Three Hundred} simulations \citep{Cui_2018} provide varying levels of agreement with observations \citep{McDonald_2017, Ghirardini_2021} of the gas profile around galaxy clusters \citep{Li_2023}.

From this brief historical excursus, it is clear that conclusions can differ depending on the astrophysical model embedded in the simulations. For a systematic study of the effects of specific stellar and AGN feedback prescriptions on the distribution of gas within haloes, it is then useful to adopt a suite of simulations with same box size, mass resolution and initial conditions, but differing only on the modelling of feedback processes. In this respect, the \simba\ suite of simulations \citep{Simba} represents an ideal test bed, incorporating six feedback variants. The suite has already been utilised to study the effect of stellar and different AGN feedback modes on the gas distribution and thermal state of gas within haloes both at high \citep{Sorini_2020} and low redshift \citep{Khrykin_2024, Yang_2024}. A more detailed analysis was provided by \cite{Sorini_2022}, who showed the evolution of the radial profile of various gaseous phases within haloes of different mass between $z=2$ and $z=0$, for five feedback variants of the \simba\ simulation. 

In this study, we build up on the previous body of work around the \simba\ simulation by undertaking a more systematic study of the evolution of the total gas density profiles. We consider haloes hosting different objects, from galaxies to clusters (mass range $10^{11} - 10^{14} \Msun$), over the redshift interval $0<z<4$. Our investigation also includes a new feedback variant of the \simba\ suite of simulations, which we present for the first time in this paper, where we do not impose a fixed cap for the maximum speed of AGN-driven jets. We find that the gas density profiles, in all runs and in the entire mass and redshift range considered, are well approximated by a power law. The slope and normalisation of the power law are sensitive to the the different feedback modes, depending on the halo mass and redshift. Thus, we argue that measuring these quantities with current and forthcoming observations has the potential to tightly constrain different feedback prescriptions. We also provide a universal fitting formula for the redshift and mass evolution of the slope and normalisation of the gas density profiles, for all runs considered. This is the main result of the paper, and can be used to mimic the effect of \simba-type feedback onto N-body simulations or semi-analytic models.

We briefly present the \simba\ simulations in \S~\ref{sec:simulations}. We analyse the gas density profiles, and the evolution of their slope and normalisation over mass and redshift, in \S~\ref{sec:profiles}. We introduce our universal fitting formula in \S~\ref{sec:fit} and discuss the applicability of limitations of our work in \S~\ref{sec:discussion}. We present our conclusions in \S~\ref{sec:conclusions}. Throughout this manuscript, unless otherwise indicated, co-moving quantities are preceded by a `c' (e.g. $\ckpc$, $\cMpc$, etc.).

\section{Simulations}
\label{sec:simulations}

\begin{table*}
\begin{center}
\caption{\simba\ runs used in this work.
}
\label{tab:runs}
\begin{tabular}{lccccccccc}
\hline
Simulation & $L_{\rm box}$ & $N$ & $m_{\rm DM}$ & $m_{\rm gas}$ & Stellar Feedback & AGN winds & AGN-jet & X-ray heating & Variable $v_{\rm jet}$ cap \\
 & ($\hMpc$) & & ($\rm M_{\odot}$) & ($\rm M_{\odot}$) & & & & & \\
\hline
Var-vjet-cap & 50 & $2\times 512^3$ & $9.6 \times 10^7$ & $1.82 \times 10^7$ & \cmark & \cmark & \cmark & \cmark & \cmark \\
{\bf Simba-50} & {\bf 50}  & $\mathbf{2\times 512^3}$ & $\mathbf{9.6 \times 10^7}$ & $\mathbf{1.82 \times 10^7}$ & \cmark & \cmark & \cmark & \cmark & --- \\
No-X-ray & 50  & $2\times 512^3$ & $9.6 \times 10^7$ & $1.82 \times 10^7$ & \cmark & \cmark & \cmark  &  --- & --- \\
No-jet & 50  & $2\times 512^3$ & $9.6 \times 10^7$ & $1.82 \times 10^7$ &\cmark & \cmark & --- & ---  & --- \\
No-AGN & 50  & $2\times 512^3$ & $9.6 \times 10^7$ & $1.82 \times 10^7$ &\cmark & --- & --- & ---  & ---  \\
No-feedback & 50  & $2\times 512^3$ & $9.6 \times 10^7$ & $1.82 \times 10^7$ & ---  & ---  & --- & --- & --- \\
\hline
Simba-100 & 100  & $2\times 1024^3$ & $9.6 \times 10^7$ & $1.82 \times 10^7$ & \cmark & \cmark & \cmark & \cmark & --- \\
Simba-25 & 25  & $2\times 256^3$ & $9.6 \times 10^7$ & $1.82 \times 10^7$ & \cmark & \cmark & \cmark & \cmark  & --- \\
Simba HighRes & 25  & $2\times 512^3$ & $1.2 \times 10^7$ & $2.28 \times 10^6$ & \cmark & \cmark & \cmark & \cmark & --- \\
\hline
\end{tabular}
\end{center}
%\raggedright
{\footnotesize From left to right, the first five columns of the table report: the name of the simulation as labelled in this manuscript; the box size; the cumulative number of initial DM particles and initial gas elements; the mass of a DM particle; the mass of a gas resolution element. The following columns display whether the specific features mentioned in the header are activated in each run (see \S~\ref{sec:simulations} for details). The fiducial run, typed in boldface, is the Simba-50 run. The main results of this work concern the comparison of the first six simulations listed in the table. The remaining three simulations are variants of the fiducial run with either different box size or mass resolution, and are used primarily for convergence testing purposes.}
\end{table*}

\subsection{Description of the simulations}
\label{sec:sim_general}

\simba\ is a suite of cosmological simulations \citep{Simba} based on the \texttt{Gizmo} hydrodynamic code \citep{Gizmo}. It represents the successor of the \mufasa\ simulations, incorporating significant enhancements such as dual-mode black hole accretion, triple-mode AGN feedback, and an on-the-fly dust evolution model. \simba\ has been found to accurately reproduce key properties of the intergalactic and circumgalactic media \citep[e.g.][]{Appleby_2020, Bradley_2022, Christiansen_2020, Yang_2022}, black holes \citep[e.g.][]{Thomas_2019, Habouzit_2021, Habouzit_2022}, and star formation history \citep[e.g.][]{Appleby_2021, Scharre_2024}. Given its widespread use in the literature, we will therefore limit the discussion of the suite of simulations to the features that are most relevant for this work, referring the interested reader to previous publications for further details \citep{Simba, Li_2019, Thomas_2019}.

\simba\ adopts the \texttt{Grackle-3.1} library \citep{Smith_2017} for radiative cooling and photoionisation heating, including metal cooling and non-equilibrium evolution of primordial elements. The evolution of the ionising UV background follows the \cite{HM12} model, adjusted for self-shielding with the \cite{Rahmati_2013} formula. Star formation is derived from the \cite{Schmidt_1959} law for $\rm H_2$, estimated using the \cite{Krumholz_2011} prescription. Gas with a hydrogen number density above $n_{\rm th} = 0.13~{\rm cm}^{-3}$ is artificially pressurised to resolve star-forming gas in the interstellar medium (ISM). Eligible ISM gas must contain $\rm H_2$ and meet specific density and temperature criteria in order to trigger star formation. The chemical enrichment model tracks eleven elements (H, He, C, N, O, Ne, Mg, Si, S, Ca, Fe) from supernovae (Type Ia and II) and AGB stars \citep{Oppenheimer_2006}.

Stellar feedback is implemented through star formation-driven galactic winds that are modelled using kinetic decoupled ejection. Such outflows represent the collective effect of Type II supernovae, radiation pressure, and stellar winds. They are implemented via a sub-grid prescription ejecting wind particles perpendicularly to their velocity and acceleration vectors. The mass loading factor and wind speed are key parameters adopted from the scaling relations found in the FIRE zoom-in simulations \citep{Muratov_2015, Angles-Alcazar_2017b}. Winds are metal-loaded, accounting for Type II supernovae yields, with 30\% of wind particles heated according to the difference between kinetic energy and the supernova energy ($u_{\rm SN} = 5.165\times 10^{15}\ {\rm erg\ g}^{-1}$), and the rest ejected at $T\approx 10^3$K. Ejected wind particles are hydrodynamically decoupled to prevent numerical inaccuracies as a result of individual gas elements having high Mach numbers compared to their environment. Additionally, cooling is disabled to allow the hot winds to transfer their thermal energy to the CGM. The wind elements recouple when either they have been for at least 2\% of the Hubble time since launch, their density is falls $0.01n_{\rm th}$, or below the ISM density and their velocity matches the one of the surrounding particles.

\simba\ includes black hole (BH) particles that accrete gas using a dual model. Non-ISM gas above $T=10^5 \K$ follows the Bondi accretion rate (`hot-accretion mode'; see \citealt{Bondi_1952}), while cooler gas within the BH kernel uses a torque-limited `cold-accretion mode', driven by gravitational instabilities \citep{Hopkins_2011, Angles-Alcazar_2013, Angles-Alcazar_2015, Angles-Alcazar_2017a}. AGN feedback is implemented with three modes, depending on the BH mass $M_{\rm BH}$ and accretion rate:
\begin{itemize}
    \item Black holes accreting rapidly ($>0.2$ times the Eddington accretion rate) eject \textit{AGN winds} characterised as purely bipolar outflows, aligned with the angular momentum of gas within the BH kernel. The speed of the AGN winds in this radiative feedback mode follows the relation, based on X-ray observations of AGN by \cite{Perna_2017a}:
    \begin{equation}
     \frac{v_{\rm AGN \, w}}{\rm km \, s^{-1}}=500+\frac{500}{3} \left(\log \frac{M_{\rm BH}}{\Msun}-6 \right) \, .
    \end{equation}
    After ejection, the winds are kinetically coupled to the nearby gas particles. AGN winds do not influence the gas temperature directly, which remains determined by the ISM pressurisation model mentioned earlier. 

    \item For BHs with a mass exceeding $10^{7.5} \, \rm M_{\odot}$, as the accretion rate in Eddington units, $f_{\rm Edd}$, drops below $0.2$, AGN feedback evolves into the \textit{AGN-jet} mode feedback. AGN jets are still implemented as purely bipolar outflows, but can achieve significantly higher speeds than the AGN radiative winds, as described by the equation below:
    \begin{equation}
      \frac{v_{\rm AGN \, jet}}{\rm km \, s^{-1}}=\frac{v_{\rm AGN \, w}}{\rm km \, s^{-1}} + 7000 \log \left( \frac{0.2}{f_{\rm Edd}} \right) \, .
    \end{equation}
    Clearly, as the accretion rate diminishes, the AGN-jet mode becomes increasingly dominant. Nevertheless, in most runs considered in this work (see Table~\ref{tab:runs} and \S~\ref{sec:sim_runs}) the velocity increase is limited to $7000 \, \rm km \, s^{-1}$ when the Eddington ratio is $f_{\rm Edd}\leq 0.02$. Instead, only in the No-vjet-cap variant, the maximum speed increase depends on the mass of the BH as in $v_{\rm cap} = 7000 (M_{\rm BH}/10^8)^{1/3}\,\rm km \, s^{-1}$. 

    \item In addition, BHs undergoing AGN jets can provide \textit{X-ray heating} feedback if the gas fraction in the host galaxy is below $0.2$. X-ray heating impacts only the gas particles within the BH kernel, and this impact is inversely proportional to the square of the distance from the BH. Inside the kernel, the temperature of the non-ISM gas rises according to the local heating flux, whereas for ISM gas, half of the X-ray energy is coupled as heating and the remaining half is transferred as kinetic energy, causing a radial outward motion in the gas particles. Through this mechanism, low-resolution ISM avoids the rapid cooling that might occur as a result of the ISM pressurisation model.

\end{itemize}

\subsection{Runs}
\label{sec:sim_runs}

The main results of this work are obtained with six variants of the \simba\ suite of hydrodynamic simulations, all starting from the same initial conditions. All boxes have the same box size ($50 \hMpc$ per side) and contain equal number of DM and gas particles ($512^3$ per species), hence they are endowed with the same mass resolution ($9.6 \times 10^7 \, \rm{M}_{\odot}$ and $1.82 \times 10^7 \, \rm{M}_{\odot}$ for DM and gas, respectively). The cosmological model underlying all simulations is consistent with Planck-16 \LCDM cosmology \citep{Planck_2016}. THe cosmological parameters are: $\Omega_{\mathrm{m}}=0.3$, $\Omega_{\Lambda}=1-\Omega_{\mathrm{m}}=0.7$, $\Omega_{\mathrm{b}}=0.048$, $h=0.68$, $\sigma_8=0.82$, $n_s=0.97$, with the usual definitions.

The Simba-50 simulation is the fiducial run, containing all features described in \S~\ref{sec:sim_general}. In four of the other five variants considered, different feedback prescriptions are deactivated (see Table~\ref{tab:runs}). Instead, the remaining run (`Var-vjet-cap') contains all stellar and AGN feedback prescriptions included in the Simba-50 simulation. However, it differs from the fiducial run for the maximum speed increment that can be achieved by the AGN-jet mode (see \S~\ref{sec:sim_general} for details). This is the new variant of the \simba\ simulation that we mentioned in \S~\ref{sec:intro} and adopt in this work for the first time.

On top of the $50 \, \hMpc$ boxes, we also consider three simulations following the same physical prescriptions as in the Simba-50 run, but with different box size or mass resolution. These additional simulations are not used for the core results of this paper, but rather for convergence testing purposes (see Appendix~\ref{app:convergence} for details).

Throughout all simulations, halos are identified using a 3D friends-of-friends algorithm incorporated in \texttt{Gizmo}, originating from V. Springel's \texttt{Gadget-3} code. A linking length set to 0.2 times the average inter-particle separation is utilised. In post-processing, we employ the \texttt{yt}-based software \textsc{Caesar} \footnote{\url{https://caesar.readthedocs.io/en/latest/}} to couple galaxies and halos. \textsc{Caesar} additionally generates a catalogue containing numerous relevant pre-computed properties of galaxies and halos. Several outcomes of this study are derived from analysing such catalogues.

\section{Gas density profiles}
\label{sec:profiles}

\begin{figure*}
    \centering
    \includegraphics[width=\textwidth]{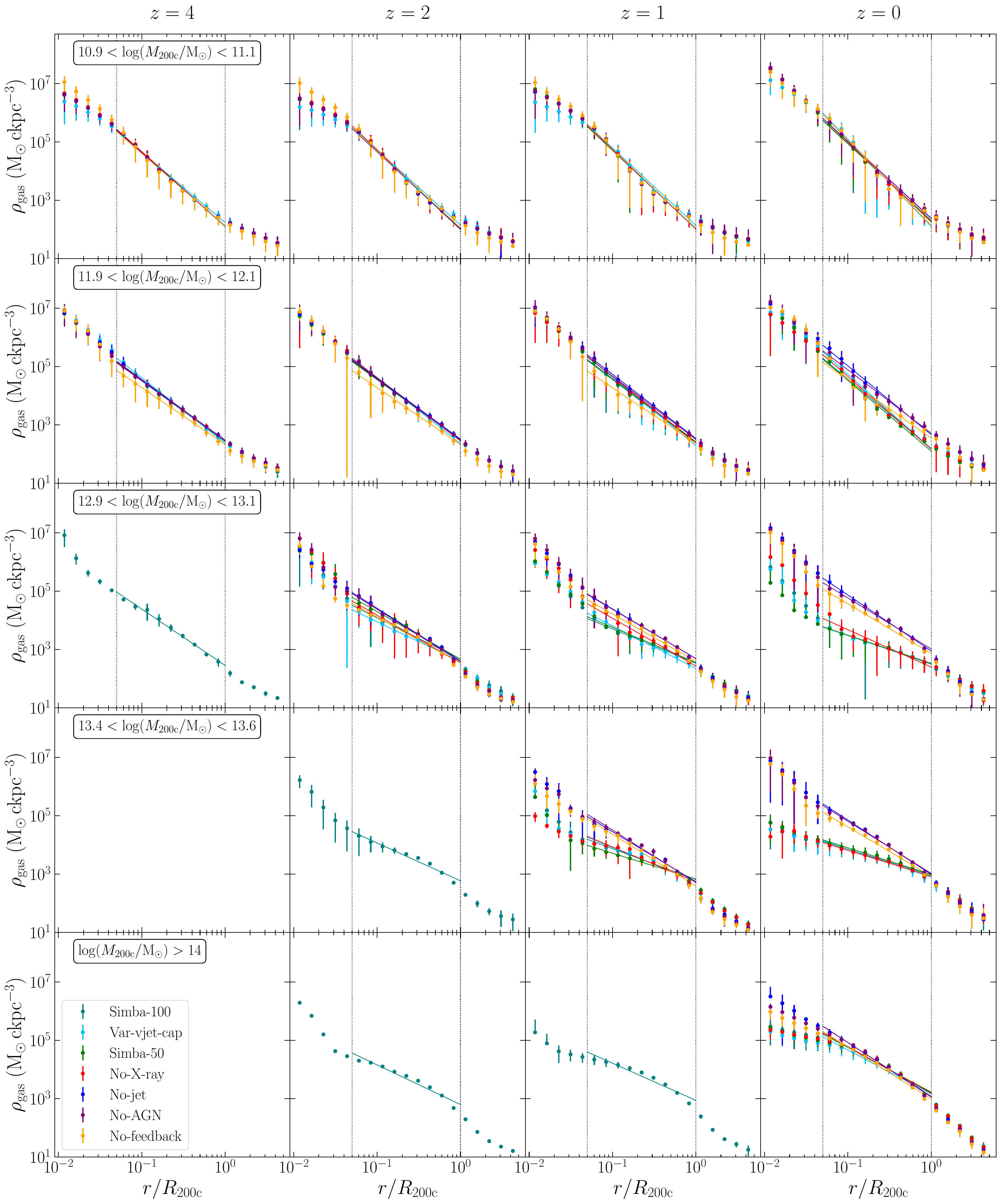}
    \caption{Co-moving density profiles of total gas within haloes of different total halo masses (as indicated in each row of panels), at different redshifts (as reported at the top of each column of panels), for the different variants of the \simba\ simulation, as specified in the legend. For each run considered, the circles represent the average density profiles of all haloes within the mass bin and redshift that correspond to the panel in question. The error bars represent the standard deviation of the gas density in each radial bin, across the haloes considered. We omit the lower error bar if its limit falls below the lowest bounds of the $x$-axis, for the sake of readability. The thin lines following the same colour coding as the markers represent the best-fit power laws to the average density profiles. The vertical dotted lines indicate the limits of radial distance within which the fits were made. Within this radial range, we find that for haloes with mass $M_{\rm 200c} \gtrsim 10^{14} \Msun$, all gas density profiles are well represented by a power law.}
    \label{fig:dens_prof}
\end{figure*}

We begin by extracting the spherically averaged radial density profiles of the gas contained within haloes in all \simba\ $50 \hMpc$ boxes, studying their dependence on halo mass and redshift. In \S~\ref{sec:dens_compare}, we will first undertake a qualitative comparison across the different feedback variants of the \simba\ suite of simulations. Then, in \S~\ref{sec:dens_fit}, we will propose a fitting formula for the gas density profiles, and qualitatively study the mass and redshift dependence of the best-fit parameters, i.e. slope and normalisation. A quantitative analysis and physical interpretation of the evolution of these parameters will follow in \S~\ref{sec:slope} and \S~\ref{sec:normalisation}.

\subsection{Qualitative comparison across different feedback runs}
\label{sec:dens_compare}

For every snapshot from $z=4$ down to $z=0$ of each feedback variant, we select all haloes wherein at least one galaxy is identified by \textsc{Caesar}. {As a result, depending on the snapshot and feedback run considered, the smallest halo selected with this criterion has a total mass between $\sim 8.1 \times 10^9 \Msun$ and $\sim 1.1 \times 10^{11} \Msun$, and contains between 72 and 1327 DM particles.

All other haloes are deemed to be unresolved for the purposes of the following analysis. For each resolved halo, we order all gas elements in 20 bins, according to their radial distance, $r$, from the position of the minimum of the gravitational potential of the halo (hereafter, `halo centre'). The first radial bin contains gas elements within $0.01 \, R_{\rm 200c}$ of the halo centre, where $R_{\rm 200c}$ is the radius enclosing a total matter density equal to 200 times the critical density of the universe, $\rho_{\rm c}$. We designate the scale $R_{\rm 200c}$ as our definition for the virial radius of the halo. The other 19 bins of radial distance span the interval $0.01 < r/R_{\rm 200c} < 5 $, with equal width in logarithmic space. We then derive the radial gas mass density profile, $\rho_{\rm gas}(r)$, by dividing the total mass of all gas elements within every bin by the volume enclosed by the spherical shells delimiting said bin.

We visualise the gas (co-moving) density profile of haloes of different mass and at different redshift from all \simba\ feedback variants in Figure~\ref{fig:dens_prof}. Throughout our analysis, we will define the total halo mass as $M_{\rm 200c}$, i.e., the mass of all matter enclosed within the virial radius $R_{\rm 200c}$ described earlier. We plot the density profiles obtained from four representative snapshots, corresponding to the redshifts indicated above each column in the figure. Within a given snapshot, we divide the haloes into four mass bins with a width of $0.2 \, \rm dex$, centred at $M_{\rm 200c}=10^{11} \Msun$, $M_{\rm 200c}=10^{12} \Msun$, $M_{\rm 200c}=10^{13} \Msun$ and $M_{\rm 200c}=10^{13.5} \Msun$, respectively. While the choice of the lower bound of the first mass bin excludes smaller haloes that passed our initial selection based on the \textsc{Caesar} halo finder, it ensures that haloes from all runs are represented at every redshift. We also include an additional halo mass bin containing all haloes above $M_{\rm 200c}> 10^{14} \Msun$. Since we are not matching haloes across different runs based on their DM mass or shared DM particles, a given halo may move between neighbouring bins when different feedback models are considered. This is not an issue in the context of this work, as we are mainly interested in understanding how the gas density profiles of a statistically significant sample of haloes in a given mass range would look under different feedback prescriptions, rather than how individual haloes would change their gas distribution when feedback mechanisms are modified. Our approach is thus more oriented towards the comparison between simulations and observations. Nevertheless, the interpretation of our results is not significantly hindered by the fact that we are not matching haloes by their unique DM particle identifiers: as shown by \cite{Sorini_2022}, the largest relative difference in the total mass of haloes selected based on their shared DM particles across different \textsc{Simba} runs is at most 25\% at $z=0$, and decreases to as little as $7\%$ at $z=4$.

Every panel in Figure~\ref{fig:dens_prof} shows the gas density profiles of haloes with a total mass comprised within the bin edges annotated inside each plot, at the redshift indicated in the upper part of the figure. We thus show the gas density profiles associated with a wide range of objects, from galaxies to groups and clusters, in the redshift interval $0<z<4$. The results obtained from different \simba\ runs are plotted with different colours, as indicated in the legend. The circles represent the arithmetic mean of the gas density at the corresponding radial bin, averaged over all haloes falling within the mass bin and snapshot considered. The vertical error bars represent the associated standard deviations. The lower error bars are omitted if their lower bound falls below minimum of the $y$-axis, to improve the readability of the figure. 

We note that at redshift $z \geq 2$, the average comoving gas density profiles are similar across all runs considered, regardless of the halo mass. However, at lower redshift ($z<2$) and within galaxy groups ($10^{13} < M_{\rm 200c}/ \mathrm{M}_{\odot} < 10^{14}$), the gas density profiles appear to decline more slowly in the runs containing at least AGN-jet feedback. Additionally, the inclusion of the AGN-jet mode reduces the normalisation of the gas density profiles. This feature suggests that the action of AGN-jet is more prominent in the redshift and mass range in question $10^{13} < M_{\rm 200c}/ \mathrm{M}_{\odot} < 10^{14}$, effectively expelling part of the gaseous component of haloes outwards. Such behaviour is consistent with the findings of previous related studies of the \simba\ suite of simulations (e.g. \citealt{Borrow_2020, Christiansen_2020, Sorini_2022, Khrykin_2024}
; see also the discussion in \citealt{Wang_He_2024}).

Within galaxy clusters ($M_{\rm 200c} > 10^{14} \Msun$), the gas density varies significantly in the inner regions of haloes ($r< 0.05 \, R_{\rm 200c}$) across the various runs. However, we verified that such radial distances typically fall below the convergence radius \citep{Power_2003, Ludlow_2019}, thus the gas density profiles within $r< 0.05 \, R_{\rm 200c}$ are not numerically reliable. On the other hand, in the region where numerical convergence is achieved ($r> 0.05 \, R_{\rm 200c}$), the gas density profiles of galaxy clusters do not exhibit as appreciable differences when the feedback prescriptions are changes. This aligns again with previous numerical studies showing that clusters approach the `closed-box approximation' \citep{Angelinelli_2022, Angelinelli_2023}, in which the baryon mass fraction is consistent with the cosmic value $f_{\rm b}=\Omega_{\rm b}/\Omega_{\rm m}$ \citep{Sorini_2022, Ayromlou_2023} and the stronger feedback heating mechanisms balance the larger amount of gas undergoing cooling \citep[e.g.][]{Allen_2011}.

There are no cluster-size haloes at $z \geq 1$ due to the limited box size of the simulations, which limits the statistics of higher-mass structures. Indeed, if we repeat our analysis for the Simba-100 run, we find that we can probe more massive haloes at every given redshift (teal data points in Figure~\ref{fig:dens_prof}). In the mass bins that are not empty in the $50 \hMpc$ boxes, the results of the Simba-100 run are consistent with those of the Simba-50 simulation. This proves that the gas density profiles of our fiducial $50 \hMpc$ run are not affected by the limited box size, even in the highest mass bins. We verified that the gas density profiles are also well converged with respect to the mass resolution of the simulations.

\subsection{Modelling the radial dependence of gas density}
\label{sec:dens_fit}

To quantitatively describe the radial dependence of the gas density, we seek a convenient fitting formula that could be applied to all feedback runs considered. The distance to which different feedback models affect the distribution of baryons outside haloes has already been quantified for the \simba\ simulation in \citealt{Sorini_2022} (see their figure 12). In this work, we will then focus on the distribution of gas within haloes, hence we will consider only data points corresponding to distance bins within the virial radius. Furthermore, we will also exclude all bins below $0.05 \, R_{\rm 200c}$, which we cannot trust due to poor numerical convergence (see \S~\ref{sec:dens_compare}). The limits of the data points that we consider for fitting the gas density profiles are delimited with vertical dotted lines in Figure~\ref{fig:dens_prof}.

We note that in the radial distance range of interest, the gas density profiles in all $50 \hMpc$ runs appear to resemble a power law, regardless of the halo mass bin and redshift. We therefore perform a minimum-$\chi^2$ fit to all profiles in Figure~\ref{fig:dens_prof}, adopting the following functional shape for the gas density:
\begin{equation}
    \label{eq:dens_fit}
    \rho_{\rm gas}(r) = \rho_R \left(\frac{r}{R_{\rm 200c}}\right)^{-\eta} \, .
\end{equation}
The free parameters $\eta$ and $\rho_R$ represent the slope and normalisation of the power law, respectively. The latter is, by definition, the value of the gas density at the virial radius, and is related to the gas mass fraction within $R_{\rm 200c}$. Equation~\eqref{eq:dens_fit} may not be an adequate fitting formula for the gas density profiles in the full range of halocentric distance plotted in Figure~\ref{fig:dens_prof}, thus our choice of restricting to the region within the virial radius will improve the accuracy of the fit. We leave the investigation of a fitting formula for the gas density profile beyond the virial radius for future work.

We now plot the resulting best-fit power laws in Figure~\ref{fig:dens_prof} with thin solid lines, using the same colour coding of the \simba\ run to which they refer. By visual inspection, the power laws provide good agreement with the data. The only exception are the gas density profiles of galaxy clusters ($M_{\rm 200c} > 10^{14} \Msun$), which exhibit a more prominent curvature, hence deviating from a power law. We rigorously confirmed our preliminary considerations by verifying that the reduced $\chi^2$ is maintained below unity for all runs, except for $M_{\rm 200c} > 10^{14} \Msun$ haloes (this aspect will be discussed further in \S~\ref{sec:discussion}). We note similar deviations from a power law in the Simba-100 run, for haloes with $M_{\rm 200c} \approx 10^{13.5} \Msun$ at $z=2$. However, there are no haloes with this mass scale in the $50 \hMpc$ boxes. Since in this work we are mainly interested in understanding the impact of feedback, we are primarily concerned with the $50 \hMpc$ \simba\ variants. Therefore, we can safely adopt the power law in equation~\eqref{eq:dens_fit} to describe the gas density profiles, as long as we exclude haloes with $M>10^{14} \Msun$ from our analysis.

Hereafter, we will then restrict ourselves to the mass range $10^{11} \lesssim M_{\rm 200c}/\mathrm{M}_{\odot} \lesssim 10^{14} \Msun$ and study the dependence of the best-fit parameters $\eta$ and $\rho_R$ on the halo mass and redshift. We begin with the slope parameter, showing the results of our analysis in Figure~\ref{fig:eta_z_rel}. We consider all available snapshots in the redshift range $0<z<4$ for all \simba\ $50 \hMpc$ runs, and organise the haloes in total mass bins of width $0.2 \, \rm dex$, centred around the values indicated at the top of every column of panels in Figure~\ref{fig:eta_z_rel}. For every mass bin, we fit the gas density profiles of each halo individually, obtaining the corresponding slope and normalisation parameters. We then split the haloes in $40 \times 40$ 2D-bins of redshift and slope $\eta$. All bins have the same width along each axis, equal to $1/40$ of the redshift interval $0<z<4$ and slope range $0<\eta<5$, respectively. For the Var-vjet-cap run, there are fewer snapshots available compared with the other runs, and some bins are empty. We plot the resulting 2D-histogram in Figure~\ref{fig:eta_z_rel}, where each row of panels refers to a different run, as indicated in the right-most panels. For any fixed redshift bin, the colour map shows the probability density function (PDF) of the slope within that bin. This is the reason why, in every run and for all mass scales, there is generally a higher density of haloes around the central region of the histogram at any fixed redshift. However, the spread in $\eta$ is substantial, especially in lower-mass haloes. 

It would certainly be useful to assess how the average and variance of the slope of the gas density profile evolves with redshift at any fixed time. To do this, for any given mass bin, we subdivide the redshift range into smaller intervals of equal width $\Delta z =0.2$ each. We then take the arithmetic mean of the gas density profile of all haloes falling within such redshift intervals, exactly as we did in \S~\ref{sec:dens_compare}. We then fit the resulting stacked density profile with the power law in equation~\eqref{eq:eta_fit}, and derive the normalisation and slope parameters. We plot the `average' slopes obtained with this procedure as black circles in Figure~\ref{fig:eta_z_rel}; the vertical bars indicate the statistical error derived from the minimum-$\chi^2$ fit, and the horizontal error bars mark the bounds of the redshift interval. The average slopes are close to the median of the PDF of $\eta$ at any fixed redshift. 

\begin{figure*}
    \centering
    \includegraphics[width=\textwidth]{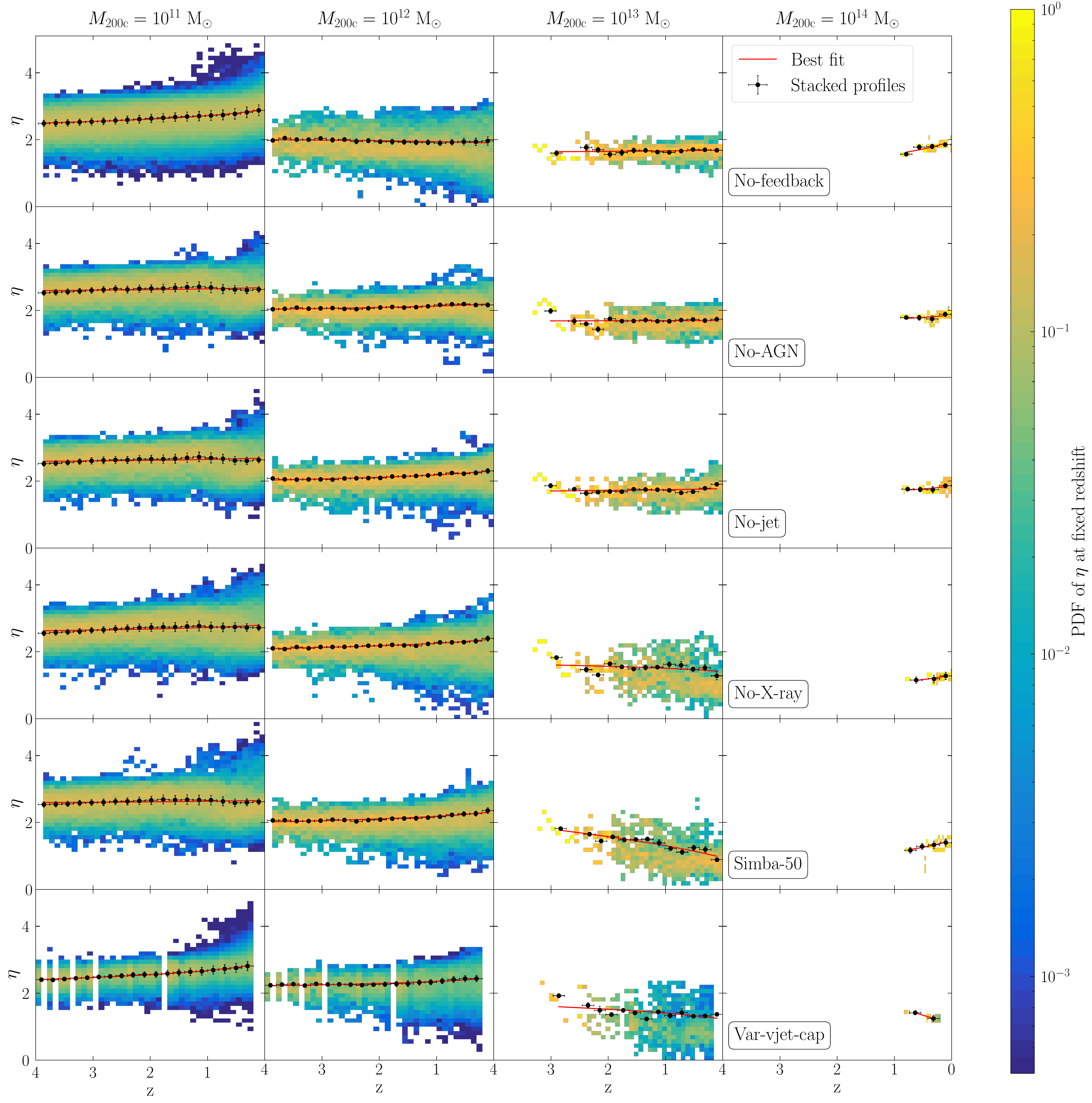}
    \caption{Redshift evolution of the slope of the gas density profile within haloes in the \simba\ $50 \, h^{-1} \cMpc$ boxes. Every row of panels refers to a different run, as specified in the left part of the figure. Each column reports the results of haloes within a total mass bin centred in the value indicated in the top part of the figure and width equal to $0.2 \, \rm dex$. The colour map in each panel shows the 2D histogram of the redshift corresponding to the snapshots considered, and the slope of the gas density profile of the haloes in the selected mass bin (see \S~\ref{sec:profiles} for details). The colour coding represents the PDF of the slope of the gas density profile at any given redshift bin. The black data points show the slope of the average gas density profile resulting from stacking all haloes in the snapshots within the redshift range indicated by the horizontal bars. The vertical error bars show the corresponding standard deviation resulting from fitting the power law in equation~\eqref{eq:dens_fit} to the average density profile. The red lines represent the best-fit power law to the slope-redshift relationship at any given total halo mass, following from equation~\eqref{eq:eta_fit}. Such fit represents an accurate description of the data at any  mass scale. For \smash{$M_{\rm 200c} < 10^{13}\Msun$}, the slope generally varies only mildly with redshift.}
    \label{fig:eta_z_rel}
\end{figure*}

\begin{figure*}
    \centering
    \includegraphics[width=\textwidth]{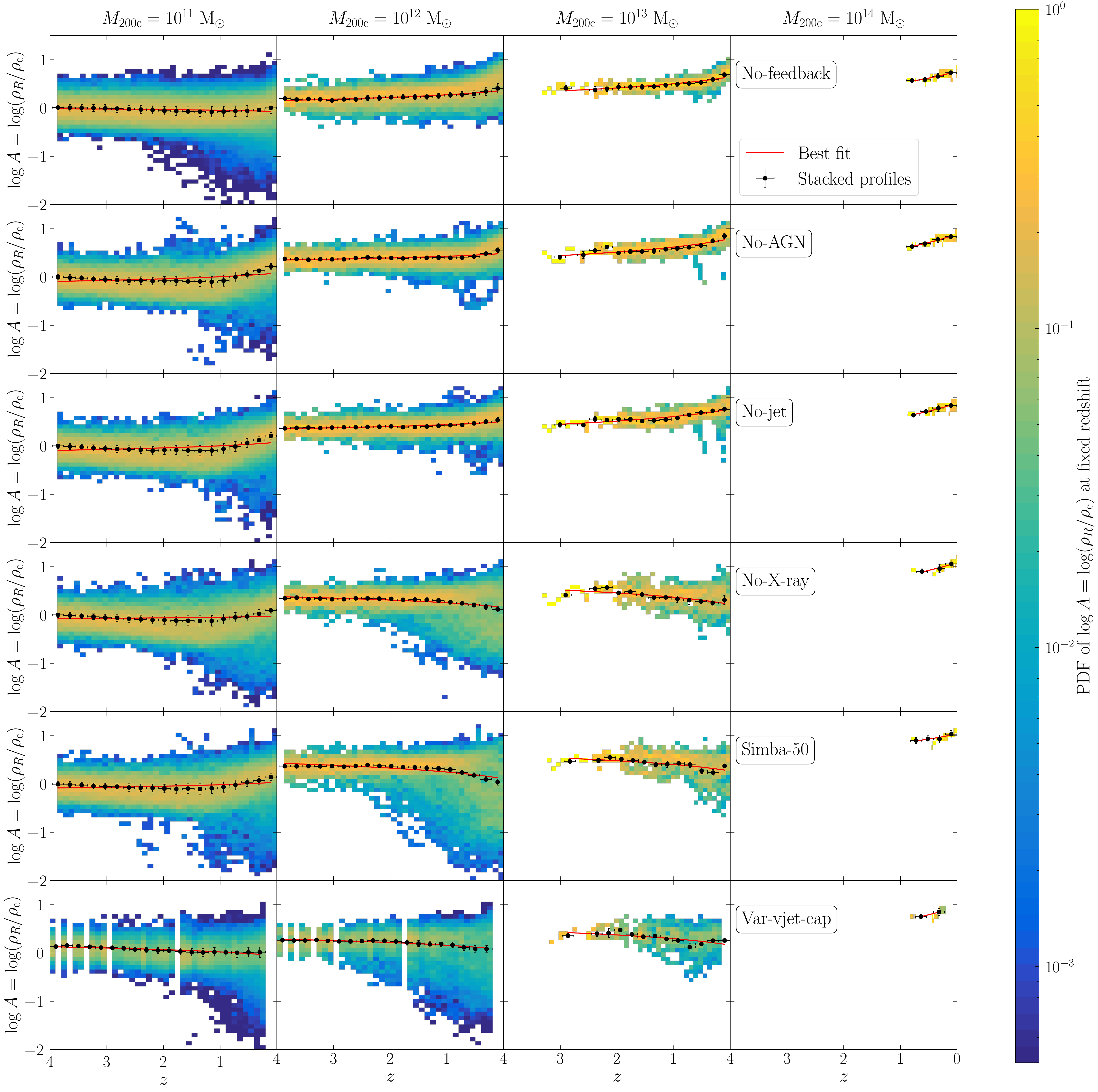}
    \caption{Redshift evolution of the normalisation of the gas density profile within haloes in the \simba\ $50 \, h^{-1} \cMpc$ boxes. Every row of panels refers to a different run, as specified in the right-most column. Each column reports the results of haloes within a total mass bin centred in the value indicated in the top part of the figure, and width equal to $0.2 \, \rm dex$. The colour map in each panel shows the 2D histogram of the redshift corresponding to the snapshots considered and the normalisation of the gas density profile of the haloes in the selected mass bin (see \S~\ref{sec:profiles} for details). The colour coding represents the PDF of the normalisation of the gas density profile at any given redshift bin. The black data points show the normalisation of the average gas density profile resulting from stacking all haloes in the snapshots within the redshift range indicated by the horizontal bars. The vertical error bars show the corresponding standard deviation resulting from fitting the power law in equation~\eqref{eq:dens_fit} to the average density profile. The red lines represent the best-fit power law to the normalisation-redshift relationship at any given total halo mass, following from equation~\eqref{eq:eta_fit}. Such fit represents a accurate description of the data at any  mass scale, although the relationship exhibits a larger scatter at $z<1$ for \smash{$M_{\rm 200c} < 10^{13} \Msun$}.}
    \label{fig:logA_z_rel}
\end{figure*}

We find a significant result: in all runs, at lower halo masses ($M \lesssim 10^{13} \Msun$), the average slope exhibits little variation with redshift. This is also true at higher masses in runs that do not include AGN-jet feedback. However, when this feedback mode is included, we observe a stronger variation of the slope of the gas density profile with redshift in galaxy groups and clusters ($M\gtrsim 10^{13} \Msun$). One might indeed expect a more conspicuous redshift evolution in higher-mass haloes, since they form more recently and, in case of major mergers, are less likely to be relaxed by present time \citep[e.g.][]{Harker_2006}. However, if the stronger evolution of the slope of the gas density profile in higher-mass haloes were connected to structure formation rather than astrophysical processes, one would expect to observe this feature in all \simba\ runs. The fact that it appears only once AGN-jet are activated strongly suggests a more direct connection between this specific feedback mode and the distribution of gas within haloes with $M_{\rm 200c} > 10^{13} \Msun$.

We now repeat the analysis described above for the normalisation parameter. Figure~\ref{fig:logA_z_rel} follows the same structure as Figure~\ref{fig:eta_z_rel}, except that the histograms are built on 2D-bins in redshift and $\log(\rho_R/\rho_{\rm c})$, which we will redefine as $\log A$ for convenience. The redshift bins are defined as before. Along the $\log A$ axis, all bins have the same width, spanning to $1/40$ of the range $-2 < \log A < 1.5$. The colour map shows the PDF of $\log A$ at any fixed redshift. We also stack the gas density profiles following the same procedure described for Figure~\ref{fig:eta_z_rel}. In this case, we derive the $\rho_R$ parameter from the minimum-$\chi^2$ fit to the stacked profiles, and then compute the corresponding `average normalisation' $\log A$, which is represented with black circles. The meaning of the error bars is analogous to Figure~\ref{fig:eta_z_rel}.

Overall, we find that the average normalisation is weakly dependent on redshift down to $z\approx 1$ at low mass ($M_{\rm 200c} \approx 10^{11} \Msun$). After $z=1$, most \simba\ runs exhibit an upturn, with the exception of the Var-vjet-cap variant. At higher masses, we observe relatively weak but qualitatively varied dependencies of the average normalisation on redshift. Generally, runs without AGN-jet exhibit a slightly increasing $\log A$, while the inclusion of this feedback mode tends to decrease the normalisation at late times. This reflects the fact that AGN-jet in \simba\ are very effective at removing gas from within higher-mass haloes, especially at later times ($z<2$; see \citealt{Sorini_2022}). As a consequence, the normalisation of the gas density profile decreases (see also Figure~\ref{fig:dens_prof}). Instead, AGN-jet are absent from $M\approx 10^{11} \Msun$, whose central BHs fail to reach the minimum mass threshold for the activation of this feedback mode \citep[see, e.g.][]{Thomas_2019, Scharre_2024}. This explains why there is no downturn of the average normalisation in the lowest mass bin showed in Figure~\ref{fig:logA_z_rel}. 

It is important to note that the spread in the normalisation--redshift relationship at any fixed mass $M_{\rm 200c} \leq 10^{12}$ is larger, in relative terms, than in the case of the slope $\eta$. This is especially true at lower redshift, where multiple feedback mechanisms are more likely to co-exist \citep{Sorini_2022}. The interplay between stellar and AGN feedback in galaxy-size haloes ($M< 10^{13} \Msun$) is notoriously complex to disentangle and can result in more varied effects on the gas distribution of individual haloes \citep[e.g.][]{Booth_2013, Delgado_2023, Tillman_2023, Gebhardt_2024}. Indeed, the spread in the normalisation--redshift relationship is larger in the bottom three rows of panels in Figure~\ref{fig:logA_z_rel}, which correspond to runs where at least AGN-jet are activated. On the other hand, the scatter around the average normalisation is smaller at higher masses $M_{\rm 200c} >10^{13} \Msun$, and particularly for galaxy clusters ($M_{\rm 200c}\approx 10^{14} \Msun$). Additionally, the normalisation--relationship is very similar for galaxy clusters across different runs. This partially follows form the fact that clusters can be approximately treated as closed boxes \citep{Angelinelli_2022, Angelinelli_2023}. The action of different feedback modes will then generally redistribute the gas density within the halo, without ejecting a relatively high gas mass fraction outside the virial radius, hence preventing the decrease in the normalisation of the gas density profile.

While feedback processes play an important role in the spread of the relationships shown in Figures~\ref{fig:eta_z_rel}-\ref{fig:logA_z_rel}, it is hard to quantify to what extent they contribute to the observed scatter, compared with other effects. Part of the variance in the slope and normalisation may follow from the intrinsic scatter in the concentration of the underlying DM density profiles, which in turn would affect the gas distribution in the halo. The fact that the scatter in both slope and normalisation seems to be comparable in all feedback runs for the lowest mass bin suggests that at least for $M_{\rm 200c} = 10^{11} \Msun$ haloes the variance in the DM structure may be the dominant source of the scatter. However, at higher masses, as discussed earlier, there is a larger spread in the redshift evolution of the normalisation parameter once the AGN-jet mode is activated, thus highlighting a more prominent role of this feedback mechanism. In any case, one should also bear in mind the potential impact of numerical effects: in principle, the limited box size may artificially decrease the scatter in the relationship at higher halo masses. Nevertheless, we verified that the normalisation is robust under different box sizes, therefore our results can be considered well converged in this respect (see Appendix~\ref{app:convergence}).

To summarise, we verified that, for all \simba\ $50 \hMpc$ simulations, the radial gas mass density profiles of haloes in the mass range $10^{11} < M_{\rm 200c}/\mathrm{M}_{\odot} < 10^{14} \Msun$ and redshift interval $0<z<4$ are well represented by a power law. For any fixed halo mass, the slope of such profiles depends only weakly on redshift for $M_{\rm 200c} \lesssim 10^{13} \Msun$, and exhibits a stronger evolution otherwise. The normalisation of the gas density profile, which is related to the gas mass fraction enclosed within the virial radius, evolves mildly at higher redshift ($z\gtrsim 2$), and more strongly at lower redshift. Such empirical trends set the stage for a more quantitative analysis aimed at capturing the overall dependence of both slope and normalisation parameters with halo mass and redshift, in all runs considered. 

\subsection{Evolution of the slope of the gas density profile}
\label{sec:slope}

\begin{figure*}
    \centering
    \includegraphics[width=0.9\textwidth]{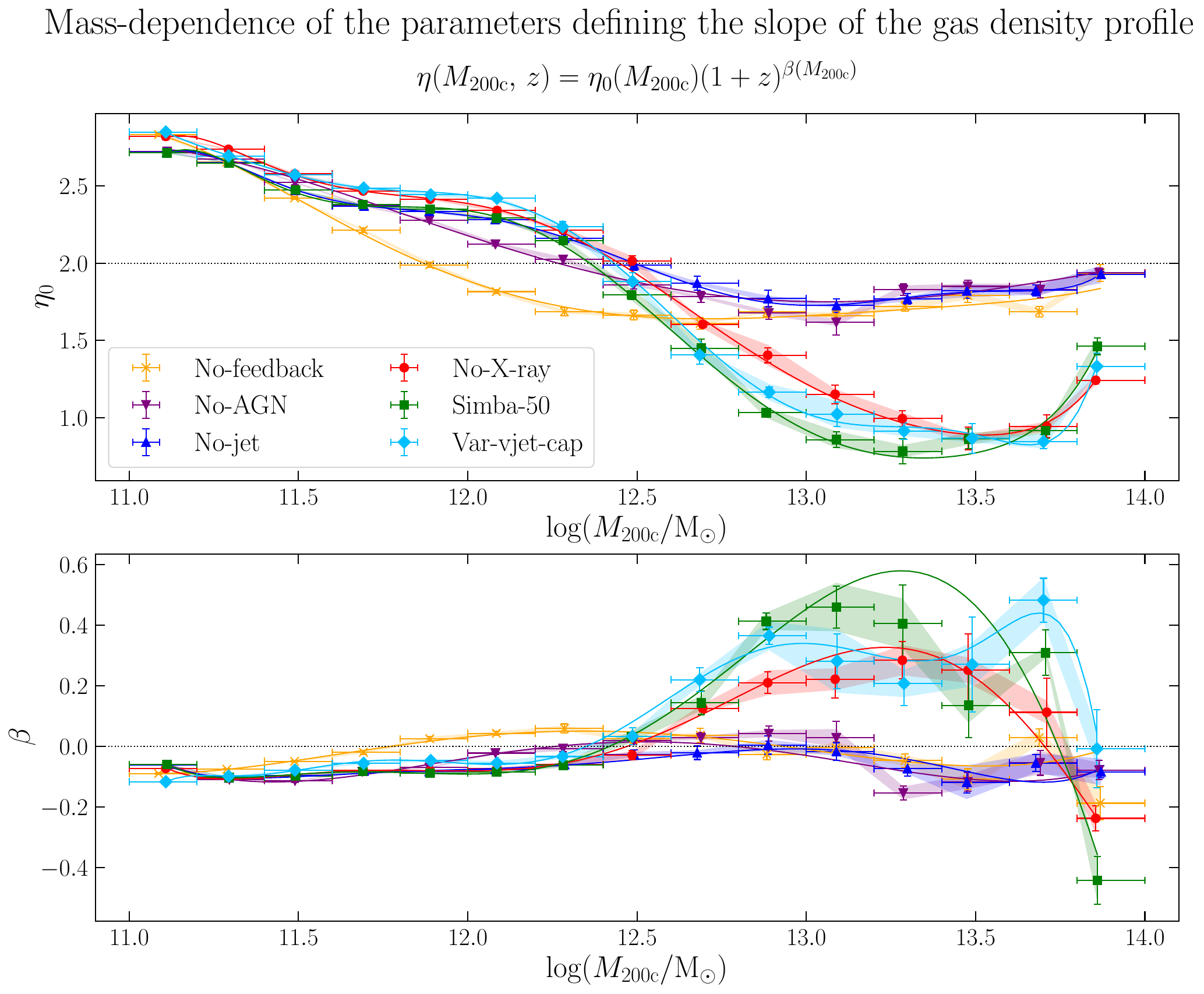}
    \caption{
    \textit{Top panel}: Mass dependence of the $\eta_0$ parameter, representing the slope of the gas density profile at redshift $z=0$ (see equation~\ref{eq:eta_fit}), for all \simba\ $50 \, h^{-1} \cMpc$ runs. The data points represent the median $M_{\rm 200c}$ in every mass bin defined by the boundaries of the horizontal bars. The vertical error bars represent the statistical error on $\eta_0$ deriving from the fit of the average gas density profile in every mass bin (see \S~\ref{sec:fit} for details), and the shaded regions show the standard deviation expected due to cosmic variance. The solid lines with the same colour coding as the data points represent the best-fit polynomials for $\eta(M_{\rm 200c})$ given by equation~\eqref{eq:eta0_fit}. The present-day slope of the gas density profile in haloes with mass \smash{$M_{\rm 200c} \gtrsim 10^{13} \Msun$} is significantly less steep once AGN jets are included in the simulation. \textit{Bottom panel}: Mass dependence of the power-law index $\beta$ in the redshift evolution of the gas density slope (see equation~\ref{eq:eta_fit}), for all \simba\ \smash{$50 \, h^{-1} \cMpc$} considered. The solid lines with the same colour coding as the data points represent the best-fit polynomials for $\beta(M_{\rm 200c})$ given by equation~\eqref{eq:beta_fit}. For haloes with mass \smash{$M_{\rm 200c} \lesssim 10^{12.5} \Msun$}, the value of $\beta$ is close to zero, thus the slope of the gas density profiles varies only mildly with redshift. Simulations that include at least the AGN-jet mode exhibit a stronger redshift evolution of the slope of the gas density profile at the higher-mass end.
    }
    \label{fig:eta_vs_M}
\end{figure*}

From Figure~\ref{fig:eta_z_rel}, we learnt that the redshift evolution of the slope of the stacked gas density profile at any fixed total halo mass is well represented by a power law. We therefore introduce the following function to describe the dependence of the average slope on mass and redshift:
\begin{equation}
\label{eq:eta_fit}
    \eta (M_{\rm 200c}, \, z) = \eta_0 (M_{\rm 200c}) \, (1+z)^{\beta (M_{\rm 200c})} \, ,
\end{equation}
where $\eta_0(M_{\rm 200c})$ and $\beta (M_{\rm 200c})$ are functions of the total halo mass, yet to be determined. At any fixed mass, $\eta_0(M_{\rm 200c})$ represents the present-day slope of the gas density profile, and $\beta (M_{\rm 200c})$ captures the rapidity of its evolution with redshift. 

To investigate the mass dependence of $\eta_0(M_{\rm 200c})$ and $\beta (M_{\rm 200c})$, we divide all gas density profiles in the redshift interval considered into logarithmic total halo mass bins with equal width ($0.2 \, \rm dex$), covering the mass range $10^{11} < M_{\rm 200c}/\mathrm{M}_{\odot} <10^{14} \, \rm M_{\odot}$. Within each bin, we obtain the slope-redshift relationship exactly as we did for Figure~\ref{fig:eta_z_rel} (see \S~\ref{sec:dens_fit} for details). We then fit the resulting relationship with equation~\eqref{eq:eta_fit} via $\chi^2$-minimisation. 

We show the values of the $\eta_0$ and $\beta$ parameters derived in every halo mass bin and for every feedback variant with different sets of data points in the upper and lower panels of Figure~\ref{fig:eta_vs_M}, respectively. The data points are plotted at the median value of the halo mass bin, which is delimited by the horizontal bars. The vertical error bars represent the statistical error on $\eta_0$ and $\beta$ derived from the fit. The shaded areas display the scatter on the parameters due to cosmic variance. This is estimated by jackknife re-sampling the haloes after splitting the simulation box of all snapshots considered into eight cubic octants of equal volume. As expected, the scatter due to cosmic variance increases at larger masses, as a consequence of the comparatively lower number of haloes. Nevertheless, the scatter is generally of the same order as the statistical error derived from the fitting procedure.

The present-day slope of the gas density profile, $\eta_0(M_{\rm 200c})$, exhibits significantly different behaviours depending on the run considered. In the No-feedback run, $\eta_0(M_{\rm 200c})$ decreases with increasing mass for $M_{\rm 200c} \lesssim 10^{12.5} \Msun$, and almost flattens at the higher-mass end around $\eta_0 \approx 1.65$. In the absence of any explicit feedback energy input, there is no outward source of pressure within the virial shock, but gas accretion following from the cooling of the CGM can still be present. The analytical cooling flow solution found by \cite{Stern_2019} predicts a slope of $\eta_0 \approx 1.6$, in excellent agreement with our findings for the No-feedback run. This does not significantly change when stellar feedback is included, as the No-AGN run exhibits a qualitatively similar trend of the present-day slope with halo mass. The main difference with respect to the No-feedback case is that, for $M_{\rm 200c} \lesssim 10^{12.5} \Msun$, the slope of the gas density profile at $z=0$ is steeper. Consequently, a slope closer to the cooling flow solution is achieved at higher masses ($M_{\rm 200c}\approx 10^{13} \Msun$) compared with the No-feedback case. This owes to the additional pressure introduced by stellar-driven winds. The situation is not significantly different once radiative AGN winds are activated. This in line with previous findings form the \simba\ suite, showing that this feedback mode only marginally affects the distribution of baryons within and outside haloes \citep{Sorini_2022, Khrykin_2024}
and the star formation history \citep{Scharre_2024}.

Once the AGN-jet mode is turned on, the present-day slope of the gas density profile becomes less steep for $M_{\rm 200c} \gtrsim 10^{12.5} \Msun$, and is heavily suppressed above $M \approx 10^{13} \Msun$, i.e., in galaxy groups. This is a consequence of the ejection power of AGN jets, which can push baryons up to $\sim 10$ virial radii from the halo (\citealt{Sorini_2022}; see also \citealt{Borrow_2020} and \citealt{Gebhardt_2024}). The energy input in \simba\ occurs via jet-inflated bubbles as observable in X-ray maps~\citep{Jennings_2024}, which over-pressurises gas at $M_{\rm 200c} \lesssim 10^{13.5} \Msun$ thereby driving gas out of the halo. The evacuation of gas results in a smoothing effect on the density profile, which then declines more gradually with the distance from the halo centre. 

However, in galaxy clusters ($M_{\rm 200c} \sim 10^{14} \Msun$) the density profile steepens up again. The addition of X-ray feedback or of a variable maximum speed for the jets does not significantly affect the trend at the higher-mass end for the \simba\ runs shown in Figure~\ref{fig:eta_vs_M}, meaning that the most important process in redistributing gas in galaxies and clusters at $z=0$ is still AGN-jet feedback. We verified that the upturn in $\eta_0$ at larger masses is dependent on the box size. For the Simba-100 simulation, the present-day slope steepens up at $M_{\rm 200c} > 10^{14} \Msun$, becoming close to $\eta_0 = 2$ at $M_{\rm 200c} \approx 10^{14.6} \Msun$ (see Appendix~\ref{app:convergence}). A slope of $\eta_0=2$ would corresponds to an isothermal gas distribution in hydrostatic equilibrium. Given that clusters are among the largest collapsed structures, their deep potential wells are able to retain halo gas despite AGN-jet feedback, and indeed in \cite{Jennings_2024} it was found that for the highest mass groups ($M_{\rm 200c} \gtrsim 10^{13.5} \Msun$) the AGN-jet energy input is at best able to balance the gas cooling but not exceed it. It would then be reasonable to expect that the resulting gas distribution is closer to hydrostatic equilibrium. However, we caution that we did not explicitly test this hypothesis in our suite of simulations.

\begin{figure*}
    \centering
    \includegraphics[width=0.9\textwidth]{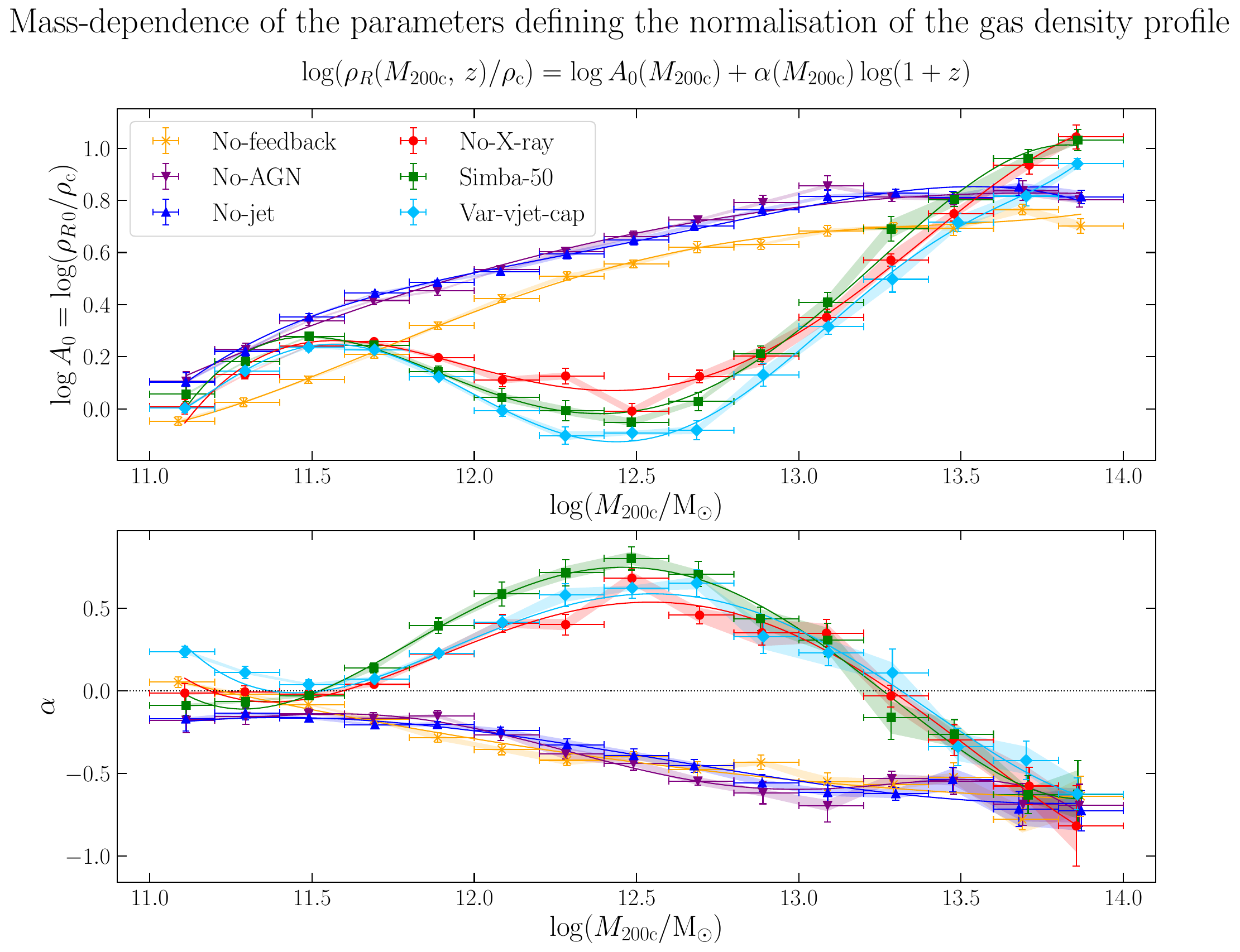}
    \caption{   
        \textit{Top panel}: Mass dependence of the $\log A_0$ parameter, representing the normalisation of the gas density profile at redshift $z=0$ (see equation~\ref{eq:logA_fit}), for all \simba\ \smash{$50 \, h^{-1} \cMpc$} runs. The data points represent the median $M_{\rm 200c}$ in every mass bin defined by the boundaries of the horizontal bars. The vertical error bars represent the statistical error on $\log A_0$ deriving from the fit of the average gas density profile in every mass bin (see \S~\ref{sec:fit} for details), and the shaded regions show the standard deviation expected due to cosmic variance. The solid lines with the same colour coding as the data points represent the best-fit polynomials for $\log A_0 (M_{\rm 200c})$ given by equation~\eqref{eq:logA0_fit}. The present-day normalisation of the gas density profile drops significantly in the total halo mass range \smash{$10^{11.4} \, \mathrm{M}_{\odot} < M_{\rm 200c} < 10^{13.5} \Msun$} with the inclusion of AGN jets in the simulation. \textit{Bottom panel}: Mass dependence of the power-law index $\alpha$ in the redshift evolution of the gas density slope (see equation~\ref{eq:logA_fit}), for all \simba\ \smash{$50 \, h^{-1} \cMpc$} considered. The solid lines with the same colour coding as the data points represent the best-fit polynomials for $\alpha(M_{\rm 200c})$ given by equation~\eqref{eq:alpha_fit}. For haloes with mass \smash{$M_{\rm 200c} \gtrsim 10^{11.5} \Msun$}, the inclusion of AGN jets imprints a qualitatively different redshift evolution of the normalisation of the gas density profile.
    }
    \label{fig:logA_vs_M}
\end{figure*}

We can understand the redshift evolution of the slope of the gas density profile from the trend of $\beta (M_{\rm 200c}$) with the halo mass. For runs where AGN jets are deactivated, we have that $\vert \beta \vert <0.12$, meaning that the slope of the gas density profiles varies only mildly with time at all mass scales. The activation of AGN jets does not change this trend for $M_{\rm 200c} \lesssim 10^{12.5} \Msun$. However, above $M_{\rm 200c} \gtrsim 10^{12.5} \Msun$ and up to $M_{\rm 200c} \approx 10^{13.6} \Msun$, we observe a larger positive index $\beta$. The slope of the gas density profile is therefore generally steeper at earlier times within galaxy groups. This reflects the fact that the ejective action of the AGN jets accumulates over time, and progressively diminishes the steepness of the gas profile. In galaxy clusters ($10^{13.8} < M_{\rm 200c} / \mathrm{M}_{\odot} < 10^{14}$) the index $\beta$ becomes either negative or consistent with zero, depending on the run considered. However, this data point is significantly affected by the limited box size (see Appendix~\ref{app:convergence}). Thus, measuring $\beta$ clusters would be particularly promising for discriminating amongst the different feedback models in galaxy groups, up until $M_{\rm 200c} < 10^{13.8} \Msun$. 

\subsection{Evolution of the normalisation of the gas density profile}
\label{sec:normalisation}

In Figure~\ref{fig:logA_z_rel} we analysed the redshift evolution of the  average logarithmic normalisation of the stacked gas density profiles for different fixed halo masses. In most cases, the redshift dependence appears to be consistent with a power law (thin red lines). However, in some of the feedback variants, the average normalisation-redshift relationship departs from a power law at late times, especially for low-mass haloes. Nevertheless, such deviations appear to be mild and, for the sake of simplicity and for consistency with our approach in \S~\ref{sec:slope}, we will again assume a power-law relationship to describe the redshift evolution of the average normalisation of the gas density profile at fixed halo mass. Whether or not this is a good approximation will be assessed at a later stage, when we will test the accuracy of our model (see \S~\ref{sec:accuracy}).

We then define the following fitting function for the normalisation:
\begin{equation}
    \label{eq:logA_fit}
    \log A (M_{\rm 200c}, \, z) = \log A_0(M_{\rm 200c})  + \alpha (M_{\rm 200c}) \log(1+z) \, ,
\end{equation}
where $\log A_0(M_{\rm 200c}) $ is the present-day value of the logarithmic normalisation $\log A \equiv \log(\rho_R / \rho_{\rm c})$ defined earlier, and $\alpha(M_{\rm 200c})$ is the power-law index expressing how fast the normalisation at a given halo mass changes with redshift. We calculate the value of $\log A_0(M_{\rm 200c})$ and $\alpha(M_{\rm 200c})$ in the same halo mass bins defined earlier for the slope of the gas density profile, using the exact same procedure described in \S~\ref{sec:slope}. The results are shown in Figure~\ref{fig:logA_vs_M}. 

The normalisation of the gas density profiles in the No-feedback run increases with the total halo mass, flattening out at $M_{\rm 200c} \gtrsim 10^{13} \Msun$. Including stellar feedback or AGN winds results in the same trend, except that the normalisation is systematically higher. Once AGN-jet feedback is activated, there is a drop in $\log A_0$ at $M_{\rm 200c} \approx 10^{11.5} \Msun$ that continues until $M_{\rm 200c} \approx 10^{12.5} \Msun$. Above this mass scale the normalisation increases again, until it plateaus at $M_{\rm 200c} \gtrsim 10^{13.5} \Msun$. 

These results can be more easily interpreted by noting that, for a fixed halo mass, the normalisation of the gas density profile is proportional to the gas mass fraction within the virial radius. Thus, the larger $\log A_0$ in the No-AGN run is explained by the larger amount of gas available following quenching due to stellar feedback. Likewise, the suppression of the normalisation in runs containing at least AGN-jet feedback reflects the lower gas mass fraction resulting from the outward displacement of gas. Analysing the mass dependence of the normalisation for a single run is somewhat less straightforward, since the proportionality between the gas mass fraction and the normalisation of the gas density profile at $z=0$ is not exact, but rather modulated by the mass-dependent present-day slope of the profile. But we verified that once this extra dependence is disentangled, then the trends observed in the upper panel of Figure~\ref{fig:logA_vs_M} are consistent with the gas mass fraction of haloes studied in previous work with the \simba\ suite \citep{Sorini_2022}.

We caution that the estimates of $\log A$ are not well converged with respect to mass resolution. The Simba-HighRes run predicts a systematically higher $\log A$ by a factor of $\sim 1.6$, mostly for $M_{\rm 200c} \lesssim 10^{12.3} \Msun$ (see Appendix~\ref{app:convergence}). One may then need to rescale the results for the present-day normalisation before comparing them with observations or other models. Such rescaling would also apply to the gas mass fraction derived from the normalisation parameter of the gas density profile. The poor convergence reflects the resolution-dependence of the outflows introduced in the different feedback prescriptions, the parameters of which are not re-tuned for every run in the suite of simulations (see Appendix~\ref{app:convergence} for further details).

In the lower panel, we see that the power-law index $\alpha$ is negative for all runs without AGN-jet feedback in the entire mass range considered. This means that the normalisation of the gas density profile increases with time, suggesting that the action of feedback (where present) is not strong enough to effectively push gas towards the outskirts of the halo, or is sub-dominant with respect to gas accretion onto the halo. On the contrary, in the mass range $10^{11.5} \, \mathrm{M}_{\odot} < M_{\rm 200c} < 10^{13.5} \Msun$, the activation of AGN-driven jets inverts this trend, underscoring the power of this feedback mode at evacuating haloes of a significant fraction of their gas over time. this is again consistent with previous results \citep{Sorini_2022}. Thus, measuring the normalisation of the gas density profile can potentially tightly constrain AGN-jet feedback in the mass range $10^{11.5} \, \mathrm{M}_{\odot} \lesssim M_{\rm 200c} \lesssim 10^{13.5} \Msun$.

\section{A universal fitting formula}
\label{sec:fit}

\begin{table*}
    \centering
    \caption{Best-fit parameters of the polynomial describing the mass dependence of the present-day slope of the gas density profile $\eta_0$ and the power-law index regulating its redshift evolution, $\beta$, as defined in equations~\eqref{eq:eta_fit}-\eqref{eq:beta_fit}.}
    \label{tab:eta_params}
    \begin{tabular}{ccccccc}
        \hline
        \textbf{Parameter} & \textbf{No-feedback} & \textbf{No-AGN} & \textbf{No-jet} & \textbf{No-X-ray} & \textbf{Simba-50} & \textbf{No-vjet-cap} \\
        \hline
        $\eta_{0, \, 0}$ & $1.656 \pm 0.033$ & $1.877 \pm 0.030$ & $1.994 \pm 0.020$ & $1.928 \pm 0.079$ & $1.759 \pm 0.090$ & $1.872 \pm 0.098$ \\
        $\eta_{0, \, 1}$ & $-0.168 \pm 0.037$ & $-0.447 \pm 0.024$ & $-0.85 \pm 0.016$ & $-1.417 \pm 0.087$ & $-1.859 \pm 0.080$ & $-1.968 \pm 0.092$ \\
        $\eta_{0, \, 2}$ & $0.487 \pm 0.073$ & $0.370 \pm 0.055$ & $0.075 \pm 0.068$ & $-0.62 \pm 0.23$ & $-0.66 \pm 0.29$ & $-0.92 \pm 0.36$ \\
        $\eta_{0, \, 3}$ & $-0.256 \pm 0.067$ & $0.085 \pm 0.037$ & $1.347 \pm 0.046$ & $1.20 \pm 0.23$ & $2.10 \pm 0.22$ & $2.75 \pm 0.30$ \\
        $\eta_{0, \, 4}$ & $-0.071 \pm 0.041$ & $-0.066 \pm 0.019$ & $0.094 \pm 0.068$ & $0.47 \pm 0.19$ & $0.59 \pm 0.28$ & $1.82 \pm 0.44$ \\
        $\eta_{0, \, 5}$ & $0.084 \pm 0.021$ &  & $-1.121 \pm 0.035$ & $-0.87 \pm 0.15$ & $-1.49 \pm 0.17$ & $-2.27 \pm 0.28$ \\
        $\eta_{0, \, 6}$ & --- & --- & $-0.021 \pm 0.021$ & $-0.057 \pm 0.056$ & $-0.073 \pm 0.071$ & $-1.47 \pm 0.20$ \\
        $\eta_{0, \, 7}$ & --- & --- & $0.2994 \pm 0.0068$ & $0.254 \pm 0.018$ & $0.412 \pm 0.025$ & $0.647 \pm 0.069$ \\
        $\eta_{0, \, 8}$ & --- & --- & --- & --- & --- & $0.426 \pm 0.026$ \\
        \hline
        $\beta_{0}$ & $0.046 \pm 0.014$ & $0.015 \pm 0.016$ & $-0.043 \pm 0.025$ & $0.011 \pm 0.037$ & $0.045 \pm 0.063$ & $0.090 \pm 0.092$ \\
        $\beta_{1}$ & $-0.051 \pm 0.021$ & $0.010 \pm 0.018$ & $0.100 \pm 0.027$ & $0.414 \pm 0.055$ & $0.619 \pm 0.097$ & $0.70 \pm 0.14$ \\
        $\beta_{2}$ & $-0.158 \pm 0.021$ & $-0.219 \pm 0.026$ & $0.039 \pm 0.066$ & $0.452 \pm 0.076$ & $0.71 \pm 0.13$ & $0.46 \pm 0.28$ \\
        $\beta_{3}$ & $0.042 \pm 0.023$ & $-0.008 \pm 0.024$ & $-0.142 \pm 0.065$ & $-0.256 \pm 0.11$ & $-0.282 \pm 0.19$ & $-1.577 \pm 0.41$ \\
        $\beta_{4}$ & $0.057 \pm 0.010$ & $0.0918 \pm 0.0097$ & $-0.161 \pm 0.047$ & $-0.517 \pm 0.063$ & $-0.69 \pm 0.11$ & $-1.31 \pm 0.28$ \\
        $\beta_{5}$ & --- & --- & $0.047 \pm 0.033$ & $0.001 \pm 0.034$ & $-0.059 \pm 0.070$ & $1.58 \pm 0.34$ \\
        $\beta_{6}$ & --- & --- & $0.070 \pm 0.010$ & $0.12 \pm 0.00$ & $0.127 \pm 0.018$ & $1.28 \pm 0.14$ \\
        $\beta_{7}$ & --- & --- & --- & $-0.394 \pm 0.019$ & $-0.495 \pm 0.069$ & $-0.495 \pm 0.069$ \\
        $\beta_{0 \, 8}$ & --- & --- & --- & --- & --- & $-0.394 \pm 0.019$\\
        \hline
    \end{tabular}

\centering
    \caption{Best-fit parameters of the polynomial describing the mass dependence of the present-day normalisation of the gas density profile $\log A_0$ and the power-law index regulating its redshift evolution, $\alpha$, as defined in equations~\eqref{eq:logA_fit}-\eqref{eq:alpha_fit}.}
    \label{tab:logA_params}
    \begin{tabular}{ccccccc}
    \hline
    \textbf{Parameter} & \textbf{No-feedback} & \textbf{No-AGN} & \textbf{No-jet} & \textbf{No-X-ray} & \textbf{Simba-50} & \textbf{No-vjet-cap} \\
    \hline
    $\log A_{0, \, 0}$ & $-0.051 \pm 0.015$ & $0.1278 \pm 0.0099$ & $0.101 \pm 0.016$ & $-0.054 \pm 0.030$ & $0.006 \pm 0.027$ & $0.002 \pm 0.023$ \\
    $\log A_{0, \, 1}$ & $0.313 \pm 0.079$ & $0.536 \pm 0.028$ & $0.892 \pm 0.091$ & $1.63 \pm 0.16$ & $1.67 \pm 0.14$ & $0.77 \pm 0.18$ \\
    $\log A_{0, \, 2}$ & $0.36 \pm 0.14$ & $-0.10 \pm 0.017$ & $-0.75 \pm 0.17$ & $-2.65 \pm 0.27$ & $-3.19 \pm 0.23$ & $0.80 \pm 0.54$ \\
    $\log A_{0, \, 3}$ & $-0.252 \pm 0.086$ & ---& $0.37 \pm 0.12$ & $1.43 \pm 0.16$ & $1.87 \pm 0.14$ & $-4.47 \pm 0.76$ \\
    $\log A_{0, \, 4}$ & $0.043 \pm 0.016$ & --- & $-0.067 \pm 0.028$ & $-0.228 \pm 0.034$ & $-0.322 \pm 0.029$ & $4.17 \pm 0.50$ \\
    $\log A_{0, \, 5}$ & --- & --- & --- & --- & --- & $-1.46 \pm 0.13$ \\
    $\log A_{0, \, 6}$ & --- & --- & --- & --- & --- & $0.178 \pm 0.010$ \\
    \hline
    $\alpha_{0}$ & $0.056 \pm 0.022$ & $-0.173 \pm 0.043$ & $-0.187 \pm 0.027$ & $0.075 \pm 0.074$ & $-0.025 \pm 0.050$ & $0.275 \pm 0.053$ \\
    $\alpha_{1}$ & $-0.440 \pm 0.053$ & $-0.05 \pm 0.29$ & $0.18 \pm 0.10$ & $-1.19 \pm 0.37$ & $-1.05 \pm 0.25$ & $-1.91 \pm 0.26$ \\
    $\alpha_{2}$ & $0.067 \pm 0.028$ & $0.86 \pm 0.71$ & $-0.34 \pm 0.11$ & $2.83 \pm 0.59$ & $3.56 \pm 0.39$ & $3.79 \pm 0.40$ \\
    $\alpha_{3}$ & --- & $-1.65 \pm 0.75$ & $0.075 \pm 0.038$ & $-1.60 \pm 0.34$ & $-2.29 \pm 0.23$ & $-2.09 \pm 0.20$ \\
    $\alpha_{4}$ & --- & $0.89 \pm 0.33$ & --- & $0.249 \pm 0.068$ & $0.401 \pm 0.043$ & $0.335 \pm 0.037$ \\
    $\alpha_{5}$ & --- & $-0.149 \pm 0.048$ & --- & --- & --- & --- \\
    \hline
    \end{tabular}

\end{table*}

In this section, we propose a universal fitting formula to describe the gas density profiles in all feedback variants of the \simba\ simulation. We will build upon the analysis  undertaken in \S~\ref{sec:slope} and \S~\ref{sec:normalisation}, by looking for a suitable analytical function for the mass-dependence of the slope and normalisation parameters. The best-fit functions introduced in this section, coupled with equation~\eqref{eq:dens_fit}, will allow us to analytically represent the average gas density profile of any halo in the mass range $10^{11} \, \mathrm{M}_{\odot} < M_{\rm 200c} < 10^{14} \Msun$ and redshift interval $0<z<4$. 

%\vspace{1ex}
\subsection{Mass-dependence of slope and normalisation}

We begin by considering the mass and redshift evolution of the slope of the gas density profile, encapsulated in equation~\eqref{eq:eta_fit}. To make the mass dependence fully explicit, we fit the values of $\eta_0$ and $\beta$ as a function of the corresponding median halo masses in each bin with a polynomial:
\begin{align}
    \label{eq:eta0_fit}
    \eta_0  (M_{\rm 200c}) &= \sum_i \eta_{0,\,i} \left[ \log \left( \frac{M_{\rm 200c}}{M_{\rm ref}} \right) \right]^i \\
    \beta  (M_{\rm 200c}) &= \sum_i \beta_i \left[ \log \left( \frac{M_{\rm 200c}}{M_{\rm ref}} \right) \right]^i  \, ,
    \label{eq:beta_fit}
\end{align}
where $M_{\rm ref}$ is a convenient normalisation constant for the halo mass, which we arbitrarily set to $M=10^{12.5} \Msun$ (the logarithmic mid-point of the mass interval considered), and all the $\eta_{0,\, i}$ and $\beta_i$ are free parameters. The choice of a polynomial relationship is not motivated by first principles, but rather practical reasons. Since the main purpose of our analysis is to find a flexible empirical relationship for the slope of the gas density profile that can be capture the impact of astrophysical processes across different \simba\ variants, we prefer to maintain our model as simple as possible. A polynomial function certainly satisfies this requirement. 

The question now is what degree each polynomial should have. We determine this by applying the Akaike's information criterion (AIC; \citealt{Akaike_1974}) to our numerical data. This rigorous information criterion ranks the quality of various models in representing a given dataset by minimising information loss while avoiding overfitting. We find that, depending on the run, we need a polynomial of degree between 4 and 8 for both the present-day slope and the $\beta$ parameter (see Appendix~\ref{app:akaike} for details). The full list of best-fit parameters for all runs is reported in Table~\ref{tab:eta_params}. The best-fit functions for the two parameters are plotted in Figure~\ref{fig:eta_vs_M} with thin solid lines, following the same colour coding as the data points to which they refer. We stress that the AIC is based on the relative comparison among the different models, and does not provide any indication on the absolute quality of the fit, which should be assessed through other methods (see \S~\ref{sec:accuracy}). 

\begin{figure*}
    \centering
    \includegraphics[width=\textwidth]{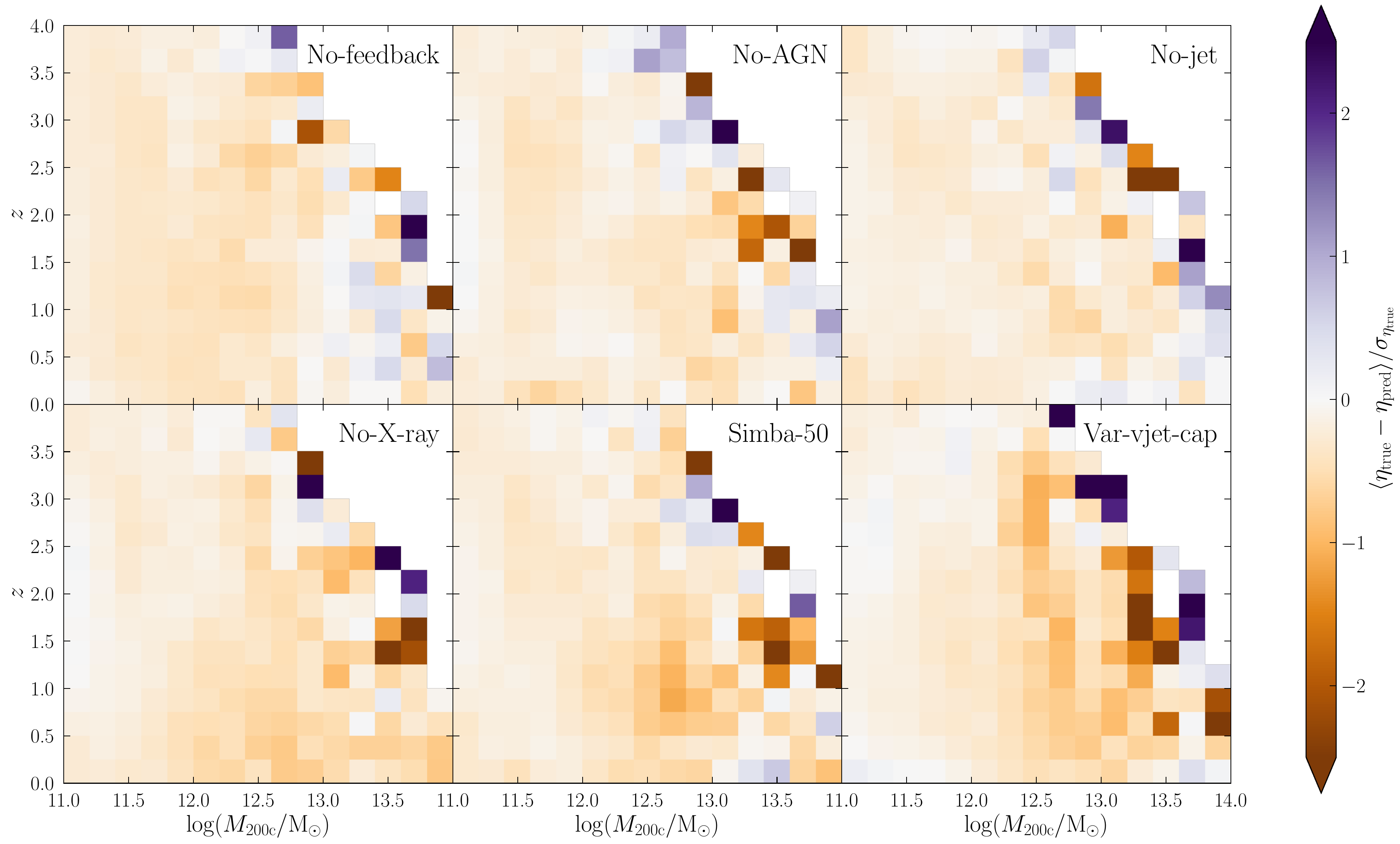}
    \includegraphics[width=\textwidth]{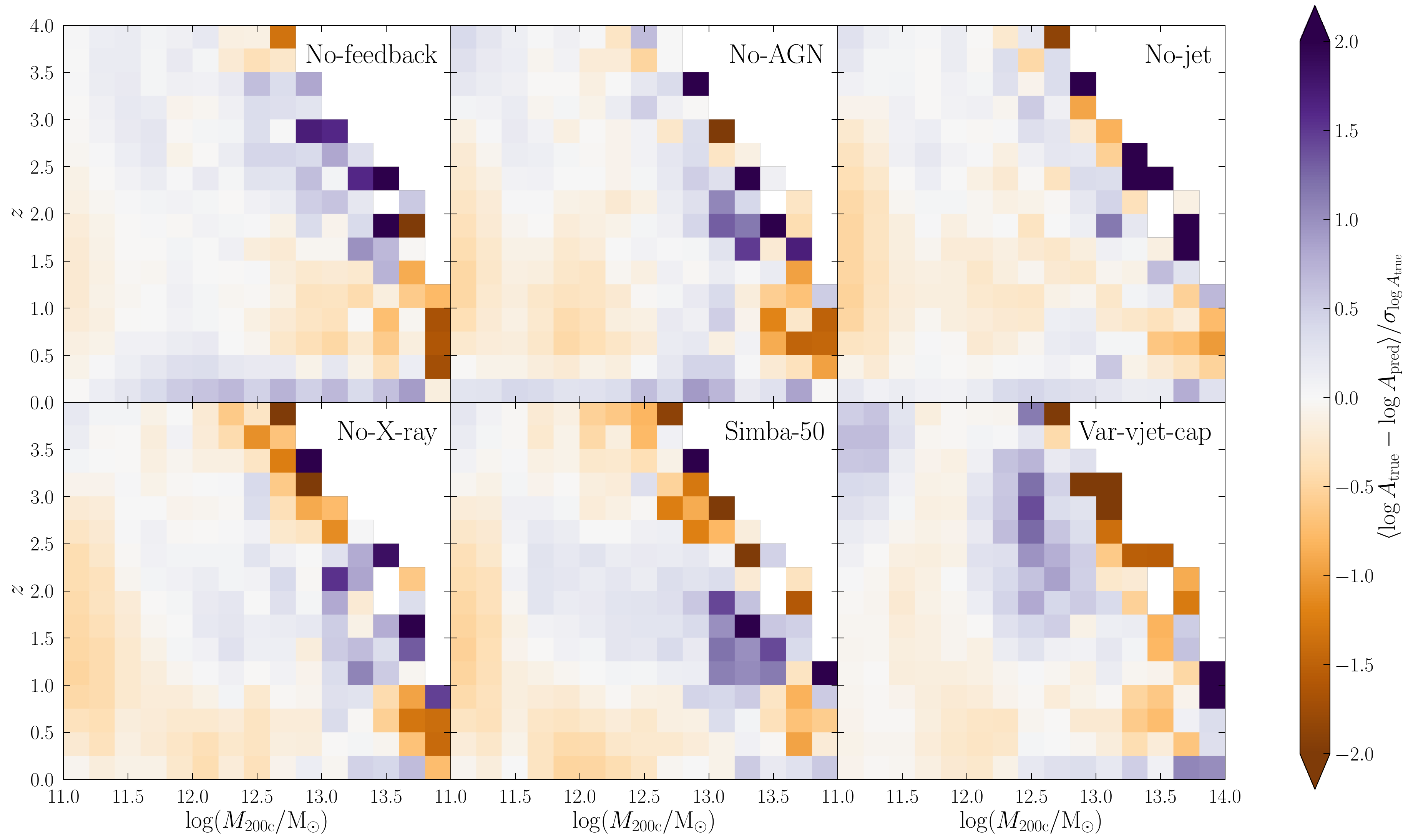}
    \caption{\textit{Top panels}: Average difference of the slope of the gas density profile of haloes within 2D-bins in redshift and total mass as measured from the simulations, and as predicted by the fitting formula given by equation~\eqref{eq:eta_fit}, for all \simba\ \smash{$50 \, h^{-1} \cMpc$} runs. The difference is normalised by the standard deviation, in each bin, of the distribution of the slope computed by the simulations. \textit{Bottom panels}: as in the upper panels, but for the predicted (equation~\ref{eq:logA_fit}) vs measured normalisation of the gas density profiles within haloes instead of the slope. As shown by the colour bar, the difference between predicted and measured slope and normalisation stays within one standard deviation for most 2D-bins for both the normalisation and slope, and exceeds two standard deviations only in some of the highest-mass bins. Our fitting formula therefore provides an accurate representation of the slope and normalisation of the gas density profile within haloes for all feedback variants of the \simba\ simulation. The fit under performs only at the boundary of the portion of the redshift-halo mass plane probed by the simulations, where the statistics of haloes is lower. }
    \label{fig:fit_accuracy}
\end{figure*}

We now undertake a similar analysis for the normalisation parameter, the evolution of which is represented by equation~\eqref{eq:logA_fit}. Having qualitatively inspected the trends of $\log A_0$ and $\alpha$ with the halo mass in \S~\ref{sec:normalisation}, we can now quantify such functions by seeking a suitable fitting function. For consistency with our study of the parameters regulating the slope-redshift-mass relationship, we choose again polynomial functions, defined as:
\begin{align}
    \label{eq:logA0_fit}
    \log A_0  (M_{\rm 200c}) &= \sum_i \log A_{0,\,i} \left[ \log \left( \frac{M_{\rm 200c}}{M_{\rm ref}} \right) \right]^i \\
    \alpha  (M_{\rm 200c}) &= \sum_i \alpha_i \left[ \log \left( \frac{M_{\rm 200c}}{M_{\rm ref}} \right) \right]^i \, ,
    \label{eq:alpha_fit}
\end{align}
where all $\log A_{0,\, i}$ and $\alpha_i$ are free parameters. We select the same pivot scale for the halo mass as in the case of the slope of the gas density profile, i..e, $M_{\rm ref}=10^{12.5} \Msun$. We determine the degree of each polynomial by applying the AIC, following the same method utilised for the slope of the gas density profile. The optimal degree is between 2 and 6 for $\log A_0$, and between 2 and 5 for $\alpha$ (see Appendix~\ref{app:akaike}). We report the values of the best-fit parameters for all runs in Table~\ref{tab:logA_params}. We remind the reader that the $\log A_0$ parameter may be underestimated due to sub-optimal numerical convergence with respect to the mass resolution, hence the corresponding fitting parameters may need to be rescaled appropriately (see also discussion in \S~\ref{sec:normalisation} and in the Appendix~\ref{app:convergence}).

At this point, all parameters needed to describe the redshift evolution and mass dependence of the slope and normalisation of the gas density profile (equations~\ref{eq:eta_fit}-\ref{eq:logA_fit}) have been specified (Tables~\ref{tab:eta_params}-\ref{tab:logA_params}). Combining equations~\ref{eq:eta0_fit}-\ref{eq:logA0_fit} with equations~\eqref{eq:dens_fit}-\eqref{eq:logA_fit}, we obtain a universal fitting formula for the average gas density profile of a halo with total mass $M_{\rm 200c}$ and redshift $z$ in any feedback variant of the \simba\ simulation.

\subsection{Accuracy of the model}
\label{sec:accuracy}

In this section, we assess the accuracy of our modelling for the evolution of the gas density profiles in the \simba\ suite of simulations. We begin by testing how the slope and normalisation of the gas density profiles predicted by our fitting formulae compare with the same quantity measured directly from the simulations.

For every feedback variant, we consider all available snapshots in the redshift range $0<z<4$, and organise the haloes respecting the resolution criteria adopted in this work in 2D-bins of halo mass and redshift. Along the mass dimension, the bins cover the range $10^{11} \, \mathrm{M}_{\odot}< M_{\rm 200c} < 10^{14} \Msun$ and have equal logarithmic width of $0.2 \, \rm dex$. The bins are linearly spaced along the redshift dimension, with width equal to $\Delta z=0.25$. We individually fit the gas density profile of all halos within every 2D-bin with the power law in equation~\eqref{eq:dens_fit}, and extract their slope, $\eta_{\rm true}$, and normalisation, $\log A_{\rm true}$. For each halo, we then calculate the the same quantities as predicted by equations~\eqref{eq:eta_fit}-\eqref{eq:logA0_fit}, $\eta_{\rm pred}$ and $\log A_{\rm pred}$, using as input for the independent variables the halo mass and redshift of the halo in question. 

To assess the accuracy of our fitting formulae for the slope, we take the arithmetic mean of the difference between $\eta_{\rm true}$ and $\eta_{\rm pred}$ in every 2D-bin in halo mass and redshift. We then normalise this average difference by the standard deviation of the slope of all haloes as measured in the simulation, $\sigma_{\eta_{\rm true}}$. This will tell us how the average accuracy of the predicted slope compares with the intrinsic scatter of the actual slopes of the gas density profiles of the haloes in every mass and redshift bin. If they are of the same order, it means that our model does a good job at reproducing the numerical data, since it does not significantly increase the dispersion in the slopes. We adopt the same procedure for determining the average accuracy of the normalisation parameter $\log A$. 

Our results for the slope and normalisation parameters are shown in the upper and lower sets of six panels, respectively, in Figure~\ref{fig:fit_accuracy}. Each panel reports the results of a different run. The squares inside every panel correspond to the 2D-bins in halo mass and redshift described earlier. The bins are colour coded according to the average accuracy at reproducing the slope (upper panels) or normalisation (lower panels) parameters, normalised by the intrinsic scatter.

\begin{figure*}
    \centering
    \includegraphics[width=\textwidth]{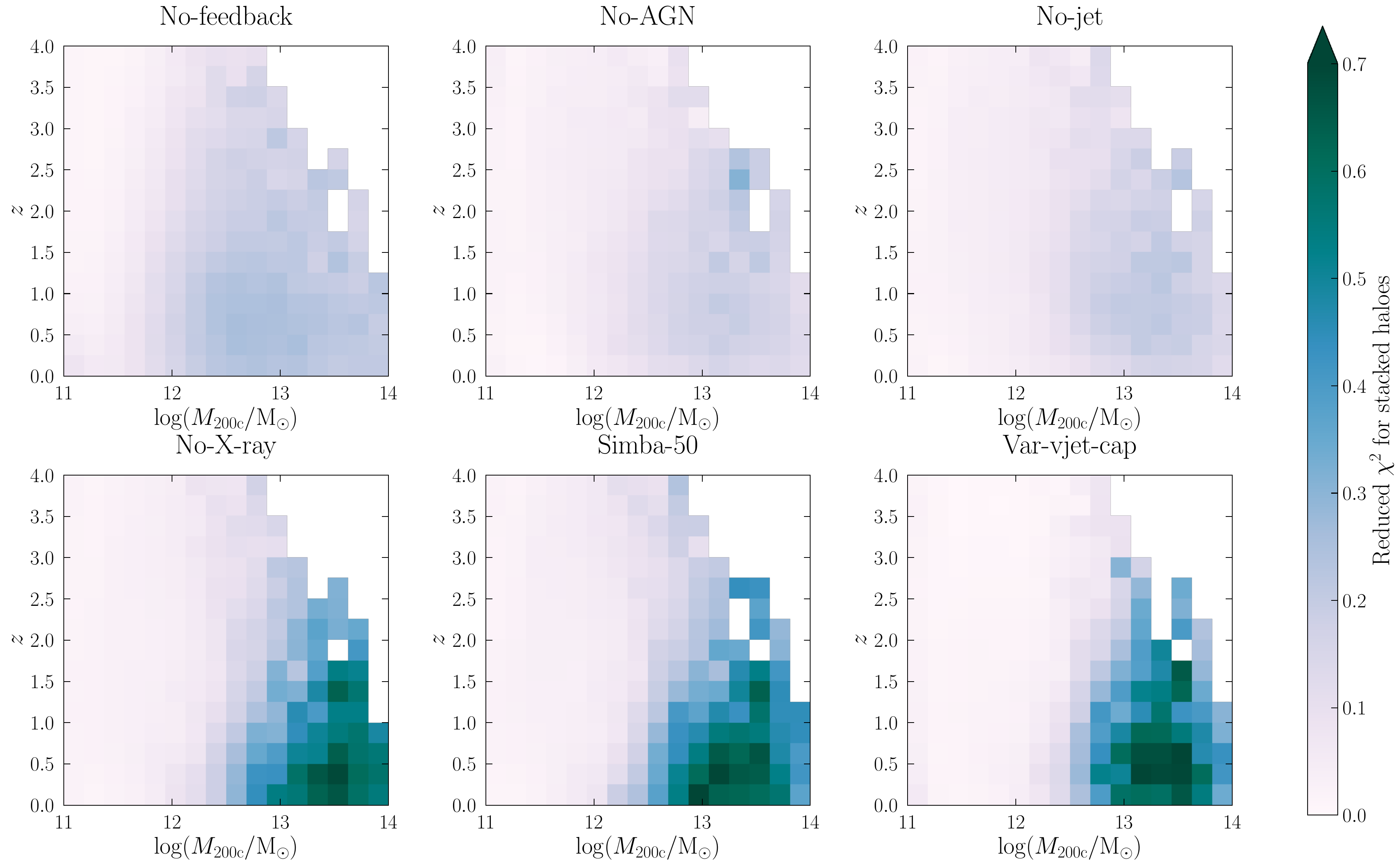}
    \caption{Average reduced $\chi^2$ obtained from a goodness-of-fit test comparing the average gas density profile resulting from stacking all haloes within different 2D-bins in redshift and total mass, with the power-law profile predicted by our fitting formulae (equation~\ref{eq:dens_fit}, combined with equations~\ref{eq:eta_fit}-\ref{eq:logA0_fit}). The low reduced $\chi^2 < 1$ values prove that our fitting formulae provide an accurate description of the gas density profiles of haloes in all \simba\ \smash{$50 \, h^{-1} \cMpc$} runs in the redshift and mass range considered.}
    \label{fig:chi_square}
\end{figure*}

For both parameters, the average accuracy is typically equal to or smaller than the intrinsic standard deviation. At the higher-mass end in all runs, the average accuracy gets generally worse, becoming comparable to $\sim 2$ standard deviations. At higher redshift ($z>1.5$), the error associated with our fitting formula in the highest-mass bins is approximately $\sim 2.5$ and $3$ sigma deviations for the normalisation and slope parameters, respectively. But these are isolated cases, affected by the low statistics of haloes in the bins in question. Indeed, in the great majority of the bins considered, our model does a remarkable job at reproducing the average slope and normalisation parameters.

We stress that the accuracy test that we performed is stricter compared to the strategy adopted for the calibration of the parameters in our fitting formulae. Indeed, the free parameters in equations~\eqref{eq:eta_fit}-\eqref{eq:logA0_fit} are obtained by fitting the \textit{stacked} gas density profiles in different redshift and halo mass bins. In Figure~\ref{fig:fit_accuracy}, we tested the accuracy of the predicted slope and normalisation by comparing them to the ones obtained by fitting \textit{individual} haloes. Therefore, our model passed a stringent test, widening the scope for its applicability (see \S~\ref{sec:discussion}).

If we perform a less strict test, following more closely the methodology utilised to derive the fitting formula for the gas density profiles, we unexpectedly obtain even better results. We divide the haloes in the same 2D-bins of redshift and halo mass adopted in Figure~\ref{fig:fit_accuracy}, and compute the average gas density profile following the same stacking procedure explained in \S~\ref{sec:dens_compare}. We then compute the gas density profile predicted by our universal fitting formula, i.e. combining equation~\eqref{eq:dens_fit} with equations~\eqref{eq:eta_fit}-\eqref{eq:logA0_fit}. The reduced $\chi^2$ resulting from a goodness-of-fit test between the stacked and predicted gas density profiles in each redshift-mass 2D-bin and every feedback variant is shown in Figure~\ref{fig:chi_square}. 

We find an excellent reduced $\chi^2$ in all 2D-bins for the runs without AGN-jet feedback. Once AGN-jet are introduced, the quality of the fit worsens at the higher-mass end for $z<2$. This is indeed the halo mass and redshift regime where AGN-jet become the dominant process in shaping the gas distribution within haloes \citep{Sorini_2022}. Nevertheless, the reduced $\chi^2$ is still within unity for all bins. 

To summarise, our universal fitting formula does an excellent job at reproducing the average gas density profiles in all \simba\ $50 \hMpc$ variants. It also correctly predicts the slope and normalisation of the gas density profiles of individual haloes with an accuracy of the same order of the intrinsic scatter of these parameters among the haloes in the simulation.

\section{Applicability and limitations}
\label{sec:discussion}

We quantified how different feedback prescriptions affect the slope and normalisation of the gas density profiles within the \simba\ simulation. Our results highlight a distinctive feature imprinted by the AGN-jet mode on the slope of the gas density profile in galaxy groups and clusters ($M_{\rm 200c} \gtrsim 10^{13} \Msun$): the present-day slope is less steep and exhibits a stronger variation with redshift (Figure~\ref{fig:eta_vs_M}) than other feedback variants. These predictions can potentially be verified with current observations. For example, \cite{Zhang_2024} measured stacked X-ray surface brightness profiles of Milky-Way-size galaxies, groups and clusters, using data from the eROSITA mission \citep{eROSITA}. They showed that the data are well represented with a beta profile; far from the core of the halo, a beta profile can be approximated with a power law, and this would enable a more direct comparison with the predictions in Figure~\ref{fig:eta_vs_M} \citep{Sorini_2024_proc}. Indeed, the core radii inferred by \cite{Zhang_2024} for their sample of objects is in the range $5-15 \kpc$, which correspond to $0.02-0.04 \, r_{\rm 200c}$. Thus, the radial distance range over which the beta profiles can be approximated with a power law ($r>0.02-0.04 \, r_{\rm 200c}$) is compatible with the halo-centric distance range considered in this work ($0.05 < r/r_{\rm 200c} < 1$).

Observations of thermal and kinetic SZ in the gaseous environment of galaxies, groups and clusters proved to be promising at constraining the impact of different feedback prescriptions on thermal state of the gas \citep[e.g.][]{Amodeo_2021, Moser_2022, Pandey_2023}. However, most of the kinetic SZ signal comes from regions beyond the virial radius \citep{Hadzhiyska_2024}, which is outside the boundaries that we considered for our fitting formula. Thus, in future work we will extend the modelling of the gas density profiles beyond the virial radius in order to best exploit the kinetic SZ data. On the other hand, at lower masses ($10^{11.5} \, \mathrm{M}_{\odot} \lesssim M_{\rm 200c} \lesssim10^{12.5} \Msun$), the predictions for the present-day slope of the gas density profile of the No-AGN and No-feedback run may be sufficiently different to be discerned by ongoing measurements. For example, observations in the optical and infrared domain with the Euclid wide survey \citep{Euclid_WideSurvey} could provide a large enough sample of galaxies to meaningfully constrain the predicted gas density profiles. 

Recent observations of $z\approx 3$ CO emitters \citep{Pensabene_2024} and of ion absorbers such as MgII around high-redshift \lya emitters traced the amount of cool gas in the CGM of galaxies across the redshift range $z\sim 3-4$ \citep{Galbiati_2024}, which is sensitive to feedback processes. Such measurements can thus be exploited for constraining the $\alpha$ parameter regulating the redshift evolution of the normalisation of the gas density profile, which exhibits a qualitatively different mass dependence when the AGN-jet mode is included (Figure~\ref{fig:logA_vs_M}). The present-day normalisation, $\log A_0$, is also very sensitive to the inclusion of AGN-jet and, somewhat less conspicuously, stellar feedback, in the mass range $10^{11.5} \, \mathrm{M}_{\odot} \lesssim M_{\rm 200c} \lesssim10^{13} \Msun$. However, this parameter is not well converged with respect to mass resolution (see Appendix~\ref{app:convergence}). While it would still be possible to constrain $\log A_0$ with observations, one would need to account for an extra source of uncertainty due to the poor numerical convergence before drawing robust conclusions.

It is important to bear in mind that there are several intermediate passages to address before making a fair comparison between the \simba\ predictions presented in this work and the aforementioned observations. Firstly, we considered the total gas mass density profile, whereas different observations trace gas with specific temperature and chemical composition. Thus, our analysis should be restricted to the gaseous phases that can be detected by different wavelength bands. Furthermore, the connection between surface brightness and density profile needs to be carefully modelled, taking into account the characteristics of the instruments used in the observations. A fair comparison would then be most effectively undertaken by generating mock observations from the \simba\ suite of simulations. We will do this in future work.

The applicability of our work is not limited to constraining feedback prescriptions from observations. In fact, our universal fitting formula can aid certain observations. For instance, mass estimates of the circumgalactic gas of $z\sim 3$ redshift galaxies from \lya emission measurements relies on assumptions of the underlying hydrogen density profile \citep{Vidal-Garcia_2024}. Re-adapting our analysis to specific atomic species would then enable more physically motivated, albeit model-dependent, estimates of the CGM mass in this kind of studies.

From a theoretical perspective, our fitting formula can be used to mimic the effects of \simba-type feedback prescriptions onto the results of N-body simulations or semi-analytic models. Indeed, as long as halo mass and redshift are known, one can apply our model and readily obtain the gas density profiles following the desired \simba\ feedback variant, without having to run full hydrodynamic simulations \textit{ab initio}. Our results can also be coupled to analytic models of cosmic star formation that rely on the simplifying assumption of spherically symmetric power-law gas density profiles within haloes \cite[e.g.][]{HS03, SP21}. Once again, our work provides a physically motivated evolution of the gas density profiles, based on sophisticated cosmological simulations.

The main limitations of this work is that our results are confined to the redshift interval $0<z<4$ and halo mass range ($10^{11} \, \mathrm{M}_{\odot} < M_{\rm 200c} < 10^{14} \Msun$). The lower mass limit is dictated by the finite mass resolution of the simulation. Running higher-resolution hydrodynamic simulations would then enable us to extend our study to dwarf galaxies and towards higher redshift, where the cutoff of the halo mass function occurs at lower masses. On the other hand, using larger boxes can extend our analysis at larger halo masses, approaching supercluster scales ($M_{\rm 200c} \sim 10^{15} \Msun$; see Appendix~\ref{app:convergence}).

Clearly, the universality of our fitting formula is still restricted to the six \simba\ variants considered. Our fitting procedure could be reasonably replicated in the runs of the CAMELS suite of simulations \citep{CAMELS} that closely follow the \simba\ galaxy formation model and extend it with many additional variations of the underlying feedback parameters \citep{Ni_2023}. However, it is not clear whether our analysis is outright portable to other cosmological simulations with significantly different feedback prescriptions. For instance, the gas density profiles in the \textsc{IllustrisTNG} \citep{Illustris_TNG_model} and \textsc{MillenniumTNG} \citep{Pakmor_2023} simulations appear to deviate appreciably from a pure power law \citep{Sorini_2024}. In these cases, either a different model for the gas density profile should be adopted, or the radial range of the profiles should be restricted such that the trend of the gas density profile is still well approximated by a power law. Indeed, we re-iterate that our model for the gas density profile is not derived from first principles. It is simply a useful empirical fitting formula that effectively captures the evolution of the gas density profiles in the different \simba\ variants, and facilitates the interpretation of the effect of feedback mechanisms on the gas distribution within haloes. A comprehensive comparison between the numerical gas density profiles from \simba\ and other simulations with the predictions of physically motivated analytical models would add more depth to this study. We leave this investigation for future work.

\section{Conclusions and perspectives}
\label{sec:conclusions}

We used several variants of the \simba\ hydrodynamic cosmological simulation to study the impact of different feedback prescriptions on the the radial gas density profiles of haloes in the total mass range $10^{11} - 10^{14} \Msun$, from $z=4$ to $z=0$. The main result of this work is a universal analytical fitting formula that accurately describes the evolution of the gas density profiles in all runs considered.

We find that at all redshifts considered, the average gas density profile at any given mass is well described by a power law within the halocentric distance range where we trust the numerical converge of our results (Figure~\ref{fig:dens_prof}). The gas distribution within haloes can thus be described with two parameters, namely the normalisation and slope of the profile. For each fixed halo mass, the redshift evolution of both the slope and normalisation parameters are also described with a power law (Figure~\ref{fig:eta_z_rel}-\ref{fig:logA_z_rel}). We thus study the mass dependence of the parameters regulating the redshift evolution of the normalisation and slope of the gas density profiles, and find that:

\begin{enumerate}
    \item When AGN-jet feedback is activated, the present-day slope of the gas density profile in galaxy groups and clusters (total halo mass $M_{\rm 200c} \gtrsim 10^{13} \, \Msun$) decreases, meaning that the gas density declines less steeply with the distance from the centre of the halo (Figure~\ref{fig:eta_vs_M}). %We argue that such distinct signature is a consequence of the effectiveness of AGN jets at .

    \item AGN-jet feedback induces a stronger redshift evolution of the slope of the gas density profile in galaxy groups and clusters, which is otherwise weak even in the presence of AGN radiative winds (Figure~\ref{fig:eta_vs_M}).

    \item The inclusion of AGN-jet feedback strongly decreases the normalisation of the gas density profile at $z=0$ in haloes with mass $10^{11.5} \, \mathrm{M}_{\odot} < M_{\rm 200c} < 10^{13.5} \Msun$, hence lowering their gas mass fraction (Figure~\ref{fig:logA_vs_M}). This follows from the evacuation of gas from haloes due to the action of AGN jets, in line with previous findings \citep[e.g.][]{Sorini_2022}. In the same halo mass range, runs that include AGN-jet exhibit an opposite trend in the redshift evolution of the normalisation of the gas density profile with respect to the \simba\ variants where this feedback mode is deactivated.

    \item Through a rigorous information criterion, we provide a universal fitting formula polynomial fitting formula for the mass dependence of the parameters describing the redshift evolution of both the slope and normalisation of the gas density profiles (equations~\ref{eq:eta_fit}-\ref{eq:alpha_fit}; Tables~\ref{tab:eta_params}-\ref{tab:logA_params}). A goodness-of-fit test of our predictions for the average gas density profile of haloes within different redshift and mass bins yields a reduced $\chi^2$ better than unity, confirming the accuracy of our model (Figure~\ref{fig:chi_square}). 

    \item Our predictions for the slope and normalisation of the gas density profiles pass an even more stringent test, where we compare them with the same quantities measured directly from the simulations, on a halo-by-halo basis. We obtain that, in all runs, the average difference between the measured and predicted quantities is consistent with the variance inherent in the sample of haloes considered (Figure~\ref{fig:fit_accuracy}). This highlights the considerable robustness of our simple fitting formulae.
    
\end{enumerate}

We provide the parameters of all best-fit functions, so that they can be straightforwardly applied to semi-analytic models, and dark-matter-only or feedback-free hydrodynamic cosmological simulations alike, in order to mimic the effect of feedback models following the \simba\ prescriptions on the gas distribution within haloes. Further work is required to verify whether other state-of-the-art cosmological simulations \citep[e.g.][]{Flamingo, Pakmor_2023} exhibit power-law gas density profiles within certain distances from the centre of haloes, or whether they favour more complex models. We will explore this issue in future work.

From the observational side, instruments such as MUSE \citep{MUSE} and surveys such as eROSITA \citep{eROSITA} recently provided high-quality data on the surface brightness profile around galaxies of different mass and redshift \citep[e.g.][]{Galbiati_2023, Galbiati_2024, Pensabene_2024, Zhang_2024}. Subject to appropriate assumptions, these measurements can be converted into corresponding gas density profiles, and can thus be utilised to constrain the predictions of cosmological simulations such as \simba. Our work serves therefore as a starting point towards this direction, which we will pursue meticulously in future work.

%----------------
% \acknowledgement
\section*{Acknowledgements}

We thank the anonymous reviewer for helpful comments. DS thanks Rachel Somerville, Shy Genel, Fred Jennings, John Peacock and the members of the \simba\ collaboration for helpful discussions, and is grateful for the support from the Post-Covid Recovery Fund of Durham University for essential travel connected to the completion of this work. This work was supported by collaborative visits funded by the Cosmology and Astroparticle Student and Postdoc Exchange Network (CASPEN). DS and SB acknowledge funding from a UK Research \& Innovation (UKRI) Future Leaders Fellowship [grant number MR/V023381/1]. RD acknowledges support from the Wolfson Research Merit Award program of the U.K. Royal Society.  DAA acknowledges support by NSF grant AST-2108944, NASA grant ATP23-0156, STScI grants JWST-GO-01712.009-A and JWST-AR-04357.001-A, Simons Foundation Award CCA-1018464, and Cottrell Scholar Award CS-CSA-2023-028 by the Research Corporation for Science Advancement. We acknowledge the \texttt{yt} team for development and support of \texttt{yt}. Throughout this work, DS was supported by the European Research Council, under grant no. 670193, by the STFC consolidated grant no. RA5496, and by the Swiss National Science Foundation (SNSF) Professorship grant no. 202671. This work used the DiRAC@Durham facility managed by the Institute for Computational Cosmology on behalf of the STFC DiRAC HPC Facility, with equipment funded by BEIS capital funding via STFC capital grants ST/K00042X/1, ST/P002293/1, ST/R002371/1 and ST/S002502/1, Durham University and STFC operations grant ST/R000832/1. DiRAC is part of the National e-Infrastructure. This work made extensive use of the NASA Astrophysics Data System and of the astro-ph preprint archive at arXiv.org. For the purpose of open access, the author has applied a Creative Commons Attribution (CC BY) licence to any Author Accepted Manuscript version arising from this submission.

%%%%%%%%%%%%%%%%%%%%%%%%%%%%%%%%%%%%%%%%%%%%%%%%%%
\section*{Data Availability}

The simulation data of all runs except for No-vjet-cap are publicly available\footnote{\url{http://simba.roe.ac.uk}}. The output of the No-vjet-cap run and the derived data underlying this article will be shared upon reasonable request to the corresponding author.

%%%%%%%%%%%%%%%%%%%%%%%%%%%%%%%%%%%%%%%%%%%%%%%%%%

%%%%%%%%%%%%%%%%%%%% REFERENCES %%%%%%%%%%%%%%%%%%

% bibiolography

\bibliographystyle{mnras}
\bibliography{halo_profiles}

%%%%%%%%%%%%%%%%%%%%%%%%%%%%%%%%%%%%%%%%%%%%%%%%%%

%%%%%%%%%%%%%%%%% APPENDICES %%%%%%%%%%%%%%%%%%%%%

\appendix

\section{Convergence tests}
\label{app:convergence}

\begin{figure*}
    \centering
    \includegraphics[width=0.85\textwidth]{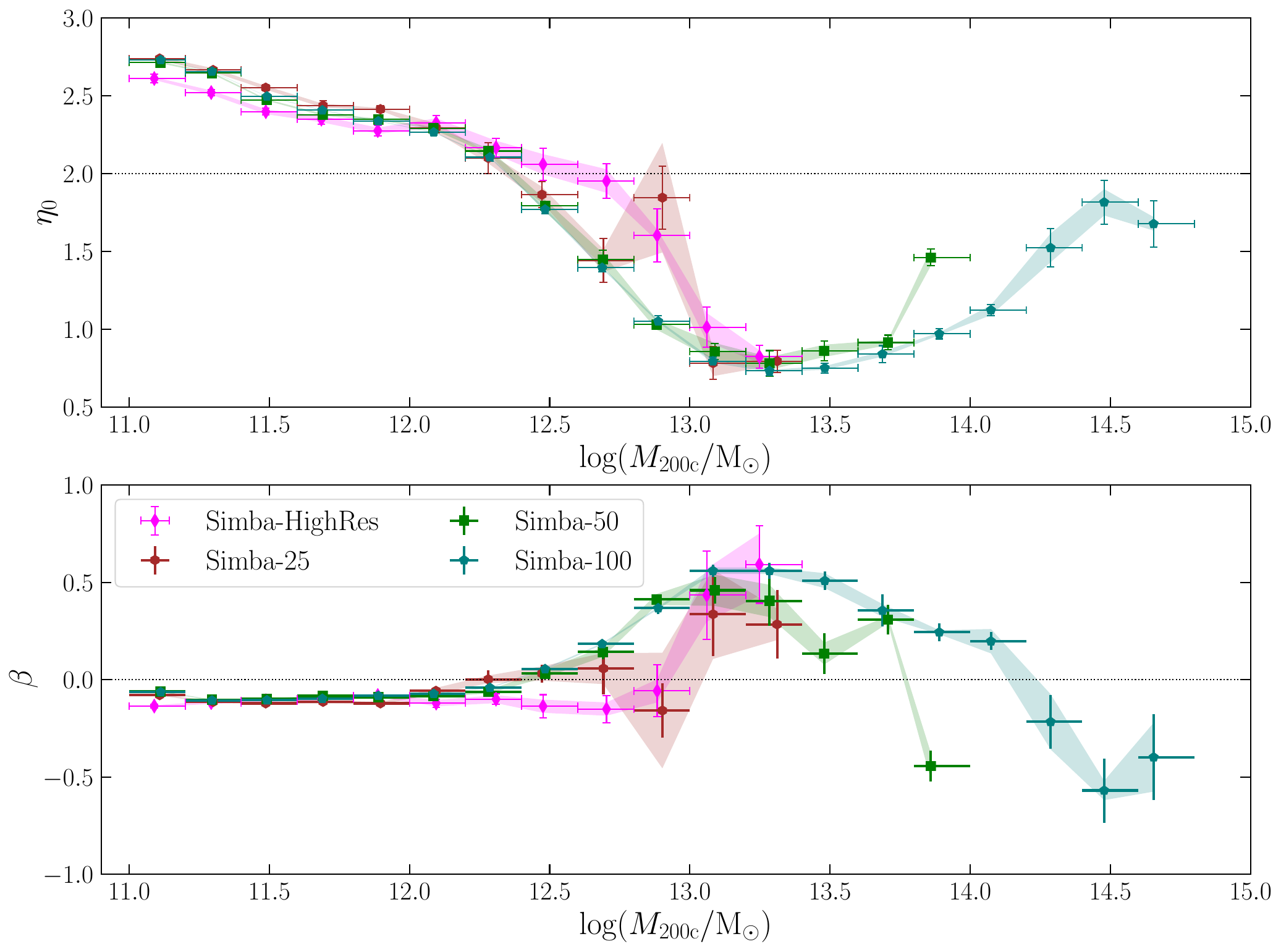}
    \includegraphics[width=0.85\textwidth]{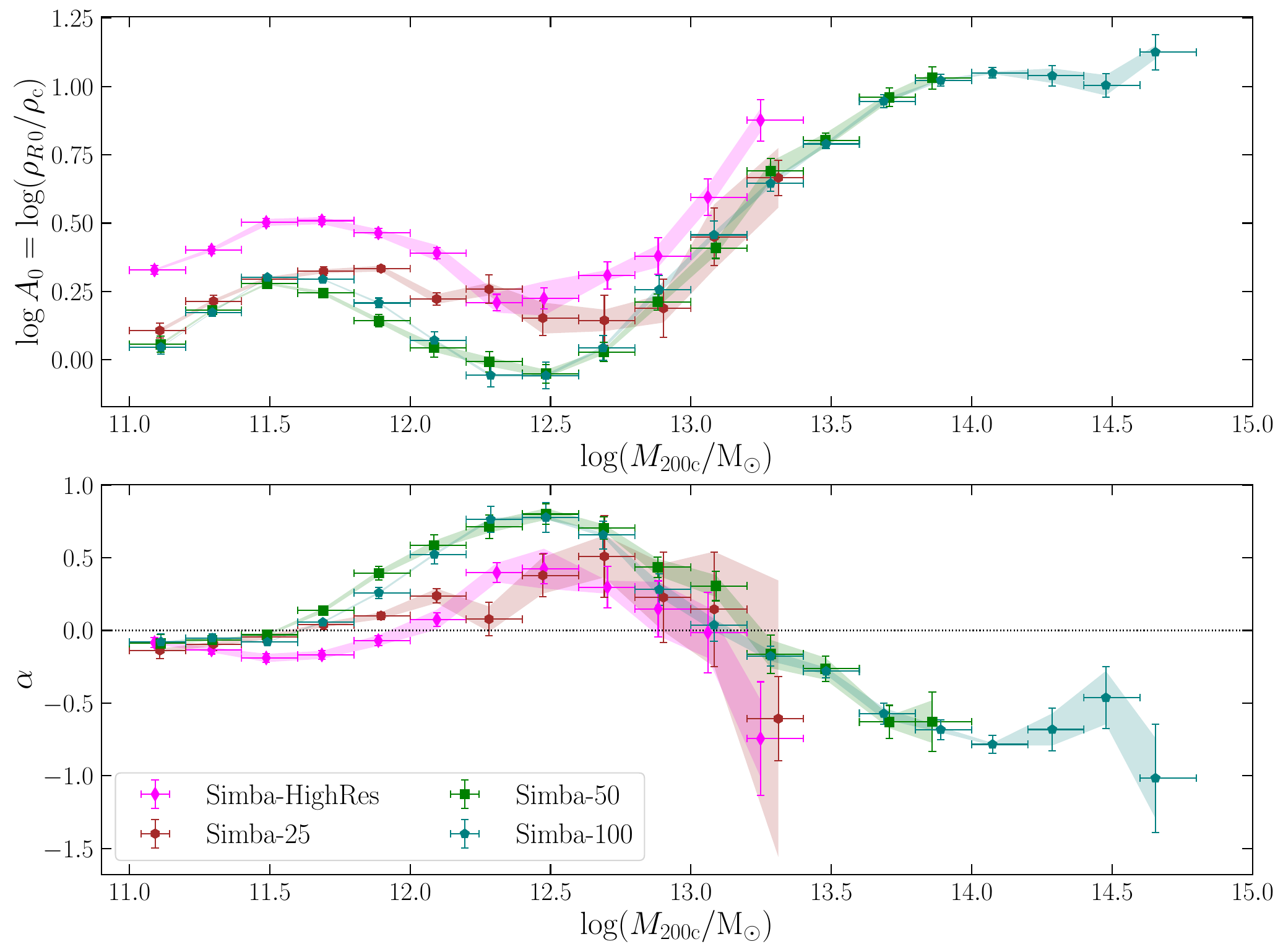}
    \caption{\textit{Top panels}: Convergence tests for the parameters of the best-fit power law to the redshift evolution of the slope of the gas density profile at any fixed total halo mass, given by equation~\eqref{eq:eta_fit}. The data points, horizontal and vertical bars, and shaded regions, have the same meaning as in Figure~\ref{fig:eta_vs_M}. The results for the parameters are well converged with respect to mass resolution, but larger boxes predict an upturn (downturn) of the $\eta_0$ ($\beta$) parameter at larger halo masses. \textit{Bottom panels}: Same as in the upper panels, but for the parameters regulating the redshift evolution of the normalisation of the slope of the gas density profile at any fixed total halo mass, as per equation~\eqref{eq:logA_fit}. The results are well converged with respect to the simulation volume and mass resolution, except that higher-resolution simulations predict a $\sim 1.6$ times larger present-day normalisation.
    }
    \label{fig:convergence}
\end{figure*}

The purpose of this section is to test the numerical convergence of the main results of our work. In Figure~\ref{fig:convergence} we analyse the halo mass dependence of the $\eta_0$, $\beta$, $\log A_0$ and $\alpha$ parameters, for the fiducial Simba-50 simulations and its variant with either different volume or mass resolution (see~\ref{tab:runs}). 

The upturn of the present-day slope of the gas density profile, $\eta_0$, at the higher-mass end occurs at a larger mass scale in the Simba-100 simulation rather than in the Simba-50 run. We find similar results for the downturn observed in the $\beta$ parameter, which regulates the redshift evolution of the slope of the gas density profile. Such features are therefore dependent on the box size of the simulation, presumably reflecting the higher statistics of cluster-size haloes that can be probed with larger volumes. Up to a halo mass of $M_{\rm 200c} \lesssim 10^{13.8}$, the Simba-50 and Simba-100 simulations provide consistent results for the present-day slope of the gas density profile. We can therefore numerically trust our results up to this mass scale. For the $\beta$ parameter, the results are fairly independent on the box size is achieved up to $M_{\rm 200c} \lesssim 10^{13.4} \Msun$. There is good convergence with respect to mass resolution over the entire halo mass range probed by the Simba-25 and Simba-HighRes boxes, since both these runs generally provide consistent results within the statistical error. 

Regarding the parameters regulating the evolution of the normalisation, $\log A_0$ and $\alpha$, the results are independent on the box size (for runs with at least $50 \hMpc$ per side, which are the ones of interest in this work) over the entire mass range considered. There is good convergence with respect to the mass resolution for the $\alpha$ parameter, whereas the two $25 \hMpc$ boxes provide statistically different results for $\log A_0$ for $M_{\rm 200c} \lesssim 10^{12.3} \Msun$. For these masses, the Simba-HighRes run exhibits an almost constant offset of $\sim 0.2 \, \rm dex$ with respect to the Simba-25 run. Thus, when predicting $\log A_0$ using our fitting formula, it might be necessary to either correct it by a factor of $\sim 1.6$, or to include the corresponding systematic error in any consideration that follows. The gas mass fraction derived from the normalisation parameter $\log A$ may thus be biased by the same factor, especially for $M_{\rm 200c} \lesssim 10^{12.3} \Msun$. We did not include any correction factor in our assessment of the accuracy of the model (Figure~\ref{fig:fit_accuracy}) because in that case we were always comparing the fitting formula derived from a $50 \hMpc$ run with the numerical data from that same run.

The sub-optimal numerical convergence is caused by the resolution-dependence of the feedback prescriptions. Since the underlying feedback parameters are not re-tuned for every run in the suite of simulations, poor convergence with respect to mass resolution can be expected for several galaxy and halo properties. At lower resolution, the conditions for outflows recoupling with the surrounding gas elements (see \S~\ref{sec:sim_general}) are typically met at larger distances from the galaxy centre. Outflows then push gas farther into the CGM and hence recycle less material due to purely numerical effects. This reduces the gas mass fraction at lower resolution, diminishing the normalisation of the gas density profile accordingly. The effect is more evident at lower halo masses because they contain fewer resolution elements.

To summarise, except for the $\log A_0$ parameter, our results are generally robust to changes in both the box size and the mass resolution over the halo mass range considered. The fitting formula for $\log A_0$ can still be utilised, as long as one bears in mind the associated offset due to the limited mass resolution.

\section{Akaike's information criterion}
\label{app:akaike}

\begin{figure*}
    \centering
    \includegraphics[width=\textwidth]{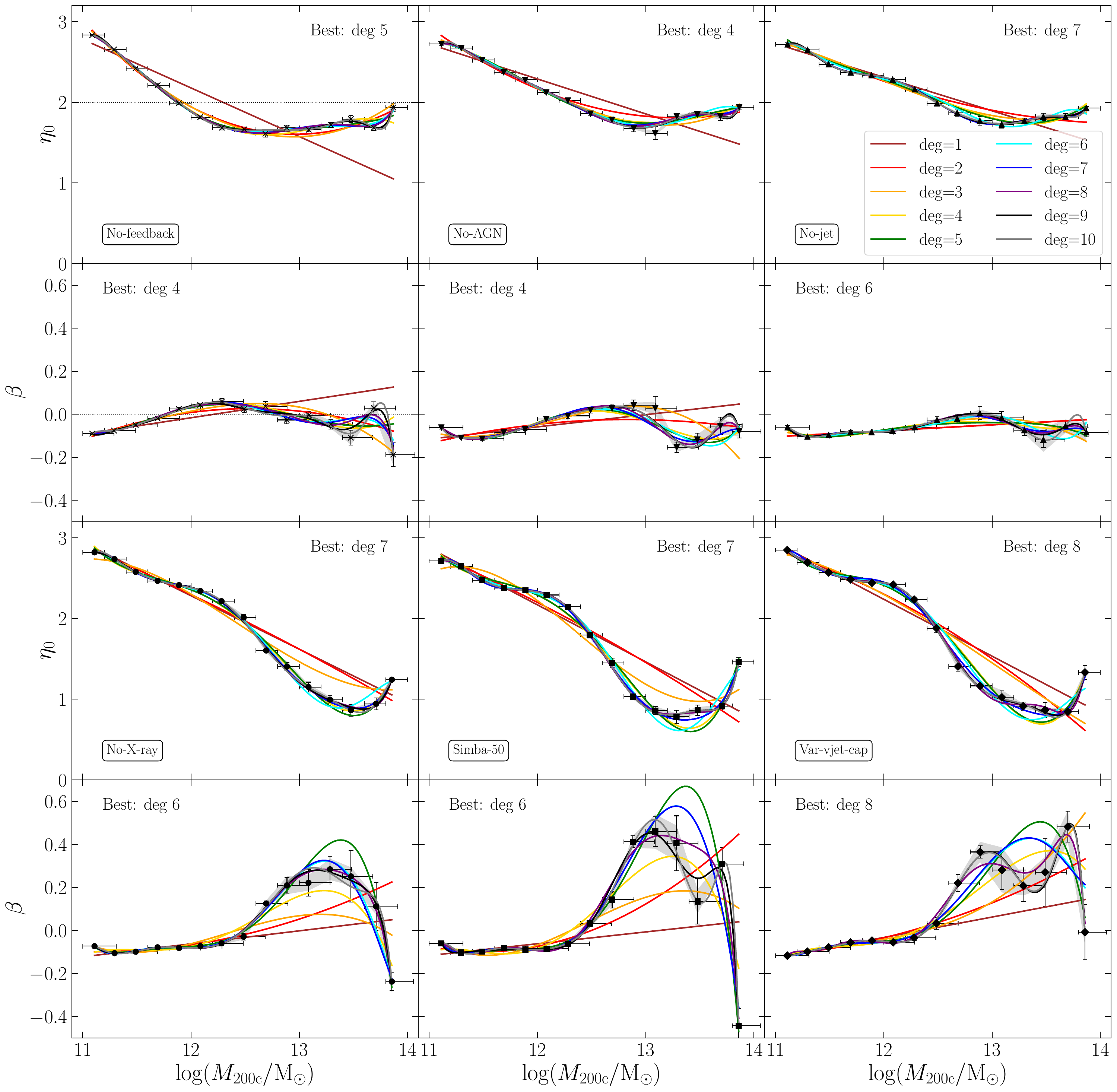}
    \caption{Results of the Akaike information criterion (AIC) test for the parameters of the power-law fit to the redshift evolution of the slope of the gas density profile at any fixed mass. For each \simba\ \smash{$50 \, h^{-1} \cMpc$} run, the upper and lower panels shows the $\eta_0$ and $\beta$ parameters, respectively, as defined in equation~\eqref{eq:eta_fit}. The black points represent the value of the parameters obtained from fitting the average gas density profile in each total halo mass bin. The points are plotted at the median halo mass in each bin, which is defined by the boundaries of the horizontal lines. The vertical error bars show the statistical error on the parameters arising from the fitting procedure, while the shaded grey region represents the scatter due to cosmic variance. In every panel, different lines correspond to polynomial fits to the data points of increasingly higher degree, following the colour coding reported in the legend inside the upper-right panel. The degree of the best polynomial fit according to the AIC test is written inside every panel.    
    }
    \label{fig:akaike_eta}
\end{figure*}

\begin{figure*}
    \centering
    \includegraphics[width=\textwidth]{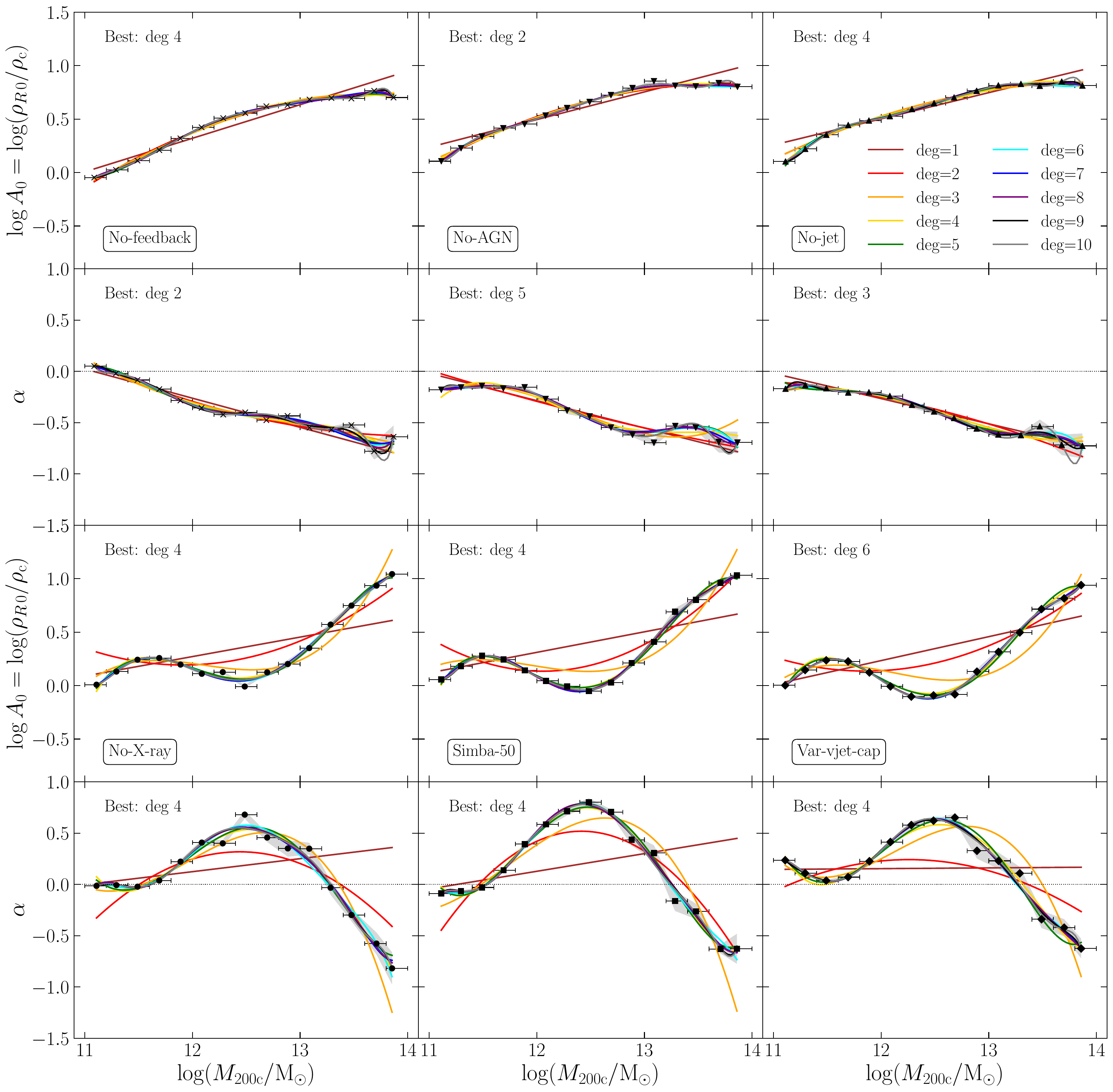}
    \caption{As in Figure~\ref{fig:akaike_eta}, but for the parameters of the power-law fit to the redshift evolution of the normalisation of the gas density profile at any fixed mass, given by equation~\eqref{eq:logA_fit}.    
    }
    \label{fig:akaike_logA}
\end{figure*}

In \S~\ref{sec:fit}, we explained that we determined the degree of the polynomials fitting the halo mass dependence of the $\eta_0$, $\beta$, $\log A_0$ and $\alpha$ parameters via Akaike's information criterion (AIC; \citealt{Akaike_1974}).

This methodology establishes an order for the quality of various models in representing a given data set, aiming to reduce the loss of information without causing overfitting. If $\widehat{\mathcal{L}}$ represents the maximised likelihood value for a specific model, and $k$ the number of free parameters, the AIC value is given by:
\begin{equation}
\label{eq:akaike}
    \mathrm{AIC} = 2 k - 2 \ln{\widehat{\mathcal{L}}} + \frac{2 k (k+1)}{n-k-1} \, ,
\end{equation}
where the final term provides an adjustment for small sample sizes $n$. The optimal model is the one with the lowest AIC value. If the smallest value among the evaluated models is $\mathrm{AIC}_{\rm min}$, then any model is $\smash{\exp[(\mathrm{AIC}- \mathrm{AIC}_{\rm min} )/2]}$ times \emph{less} likely to minimise information loss compared to the best model.

For each parameter, we derive the polynomial functions with degree $\rm 1 \leq deg \leq 10$ that provide the best fit to the numerical data. The data sets and corresponding best-fit polynomials for the parameters describing the evolution of the slope and normalisation of the gas density profiles are shown in Figure~\ref{fig:akaike_eta} and Figure~\ref{fig:akaike_logA}, respectively. For all runs, we compute the maximum likelihood for each polynomial model given the data, assuming independence of the data points. The expectation values are determined by using the fitting function on the median halo mass within each bin. For each data set, the variances are set to be the maximum between the statistical error derived from the power-law fit in each bin (via equation~\ref{eq:eta_fit} or equation~\ref{eq:logA_fit}) and the scatter due to cosmic variance. Subsequently, we substitute the maximum likelihood into equation~\eqref{eq:akaike}. The degree of the polynomial that yields the minimum AIC value is indicated inside each panel of Figures~\ref{fig:akaike_eta}-\ref{fig:akaike_logA}. These are the optimal models whose parameters are reported in Tables~\ref{tab:eta_params}-\ref{tab:logA_params}.
\vspace{2 ex}

%%%%%%%%%%%%%%%%%%%%%%%%%%%%%%%%%%%%%%%%%%%%%%%%%%

% Don't change these lines
%\bsp	% typesetting comment
%\label{lastpage}
\end{document}